\documentclass[
reprint,
superscriptaddress,
amsmath,amssymb,amsart,
aps,
prb,
]{revtex4-2}

\usepackage{amsmath}
\usepackage{amsfonts}
\usepackage{amssymb}
\usepackage{graphicx}
\usepackage{dcolumn}
\usepackage{bm}
\usepackage{accents}
\usepackage{braket}
\usepackage{amsthm}
\usepackage{hyperref}
\usepackage{bookmark} 
\usepackage{xcolor}
\usepackage{booktabs} 
\usepackage{diagbox}
\usepackage{lineno}
\usepackage{cancel}
\usepackage{orcidlink}

\graphicspath{{./figs/long}}

\begin{document}

\title{Field Theory of Linear Spin-Waves in Finite Textured Ferromagnets}

\author{T. Valet \orcidlink{0000-0002-7775-9395}}
\email[Corresponding author :]{ tvalet@mphysx.com}
\affiliation{Université Grenoble Alpes, CEA, CNRS, Grenoble INP, Spintec, 38054 Grenoble, France}

\author{K. Yamamoto \orcidlink{0000-0001-9888-4796}}
\affiliation{Advanced Science Research Center, Japan Atomic Energy Agency, Tokai, Ibaraki 319-1195, Japan}

\author{B. Pigeau \orcidlink{0000-0003-2331-8436}}
\affiliation{Université Grenoble Alpes, CNRS, Grenoble INP, Institut Néel, Grenoble, France}

\author{G. de Loubens \orcidlink{0000-0001-8096-3058}}
\affiliation{SPEC, CEA, CNRS, Université Paris-Saclay, 91191 Gif-sur-Yvette, France}

\author{O. Klein \orcidlink{0000-0001-9429-5150}}
\email[Corresponding author :]{ oklein@cea.fr}
\affiliation{Université Grenoble Alpes, CEA, CNRS, Grenoble INP, Spintec, 38054 Grenoble, France}

\begin{abstract}
In the context of an ever-expanding experimental and theoretical interest in the magnetization dynamics of mesoscopic magnetic structures, both in the classical and quantum regimes, we formulate a low energy field theory for the linear spin-waves in finite and textured ferromagnets and we perform its constrained canonical quantization. The introduction of a manifestly gauge invariant Lagrangian enables a straightforward application of the Noether's theorem. Taking advantage of this in the context of a broad class of axisymmetric ferromagnets of special conceptual and experimental relevance, a general expression of the conserved and quantized spin-wave total angular momentum is rigorously derived, while separate conservation and quantization of its orbital and spin components are established for a more restricted class of uniaxial exchange ferromagnets. Further particularizing this general framework to the case of axially saturated magnetic thin disks, we develop a semi-analytic theory of the low frequency part of the exchange-dipole azimuthal spin wave spectrum, providing a powerful theoretical platform for the analysis and interpretation of magnetic resonance experiments on magnetic microdots as further demonstrated in a joint paper~\cite{Valet2025}.      
\end{abstract}

\hfill \textbf{BC15175}
\maketitle

\section{Introduction}

Magnetization dynamics in confined mesoscopic structures is a subject of extensive ongoing research. While its potential significance for applications in memories \cite{Pigeau2010}, sensors \cite{Losby2015,Suess2018}, microwave isolators \cite{Darques2010}, and reservoir computing devices \cite{Koerber2023} has often been emphasized, it is of great interest equally from fundamental perspectives as a rare example of vector field whose expectation value does not vanish in its ground state and spontaneously breaks time-reversal symmetry. The latter property implies an intrinsic chirality of magnetic excitations, which is related to the angular momentum (AM) carried by the underlying electron spins. The chirality of ferromagnetic resonances and spin waves (SWs) has been widely utilized as a source and probe of spin angular momentum (SAM) for conduction electrons in the context of spintronics~\cite{Chumak2022}. Magnons, quanta of SWs, can potentially provide complimentary functionalities to those of photons in quantum information~\cite{zare:2022,yuan:2022}. Their theoretical descriptions so far have in some way or another invoked the notion of quantum spin operators in discrete models~\cite{holstein:1940}. Even in the recent attempts based on continuum micromagnetic models that are more appropriate in the regime of interest for quantum applications~\cite{mills:2006,serpico:2024}, the magnon commutation relation was postulated so as to agree with the discrete formulation instead of being derived from a principle. Compared to how well-recognized the notion of SAM conservation and transfer is, studies on the total AM carried by the magnetic excitations themselves have been few and far between in proportion. It is a conserved quantity of magnetization dynamics in systems invariant under rotation about a certain axis of symmetry, and can be systematically characterized by an application of Noether's theorem to the SW Lagrangian~\cite{tsukernik:1966,goldstein:1984}. A few recent works revisited the derivation motivated by the ambiguity in its interpretation~\cite{yan:2013,tchernyshyov:2015,streib:2021} stemming from the intrinsic gauge dependence of the fully non-linear Lagrangian for spin~\cite{doring:1948,miltat:2002}. They largely focus on the formal aspects and lack discussions of practical consequences in ferromagnetic resonance experiments. Specifically, as a wave field in the continuum description, magnetic excitations can carry orbital angular momentum (OAM) in addition to SAM. OAM has been a recent hot topic in photonics~\cite{franke:2022} in which it is represented by a helical wavefront about the ray of energy flux~\cite{coullet:1989,allen:1992,hancock:2021}. An analogous concept for standing waves in a confined structure has been established for phonons with rotating wavefronts~\cite{garanin:2015}. In these examples, the AM can be used as an additional label of eigenmodes \cite{Bonin2012} of the linear wave dynamics, and has attracted attentions for mode multiplexing information and parallel communication channels~\cite{chen:2020}. Essentially the same notion was known to Walker for magnetostatic SWs in uniformly magnetized spheroids~\cite{walker:1957,walker1958,Osada2018}, which have been met with a renewed interest in cavity optomagnonics~\cite{sharma:2017,osada:2018,Gloppe2019}. Given their ultimate goal being quantum technologies, consistent quantum formulation of micromagnetic AM eigenstates is a crucial unaddressed problem. The AM labeling is applicable to other axially symmetric geometries such as nanopillars~\cite{Bunyaev2015} and disks~\cite{Wang2022,Kamenetskii2024}, as well as spatially nonuniform equilibrium states including magnetic vortex~\cite{Ivanov1998,Uzunova2023} and skyrmion~\cite{Mruczkiewicz2017,Mruczkiewicz2018,Lobanov2024}. While the properties of SW eigenmodes in a vortex background have been studied in details~\cite{Ivanov1998,Buess2005,guslienko:2008,Vogt2011,Taurel2016,Schultheiss:2019,Verba2021,Mayr2021}, their connection to OAM appears to have escaped the attention. Concrete applications of the theory of SW AM have so far been limited to extended cylindrical structures~\cite{rychly2018,jia:2019,jiang:2020,Hussain2021,Korber2021,lee:2022,Korber2023}, and AM of confined standing SWs remains unexplored.

In this paper, we formulate the linear SW dynamics in finite ferromagnets over a general textured background magnetization as a classical field theory, and considering the sole continuous global spatial symmetry that may survive {\em i.e.}, the invariance under infinitesimal rotation, we derive general expressions for the associated conserved angular momenta with the help of the Noether's theorem (sec. II). The canonical quantization of the theory is rigorously performed thanks to Dirac's approach applicable to constrained systems, identifying the eigenvalues of the infinitesimal generators of rotation with the angular momenta quantum numbers (sec. III). We identify the general conditions satisfied by axisymmetric ferromagnets and allowing for the conservation of the total AM, while more restrictive conditions define a narrower class of uniaxial exchange ferromagnets in which the SAM and OAM are independently conserved. In both case the admissible general forms of the SW eigenfunctions are derived (sec. IV). In a final part (sec. V), we focus on a special type of finite systems of special conceptual and experimental relevance {\em i.e.}, axially saturated ferromagnetic cylindrical microdots, and we supplement the general formalism with a semi-analytical theory of the azimuthal exchange-dipole SWs enabling an unambiguous definition of the dynamical dipole-dipole interaction (DDI) induced spin-orbit interaction (SOI), a subject that we further investigate in connection with experimental evidences in a joint paper~\cite{Valet2025}. In conclusion, we summarize our main findings, and open some perspectives towards further development of our general theoretical framework and semi-analytical methods.

\section{Long wavelength linear spin-wave dynamics as a classical field theory}

\begin{figure}[ht!]
\includegraphics[width=0.45\textwidth]{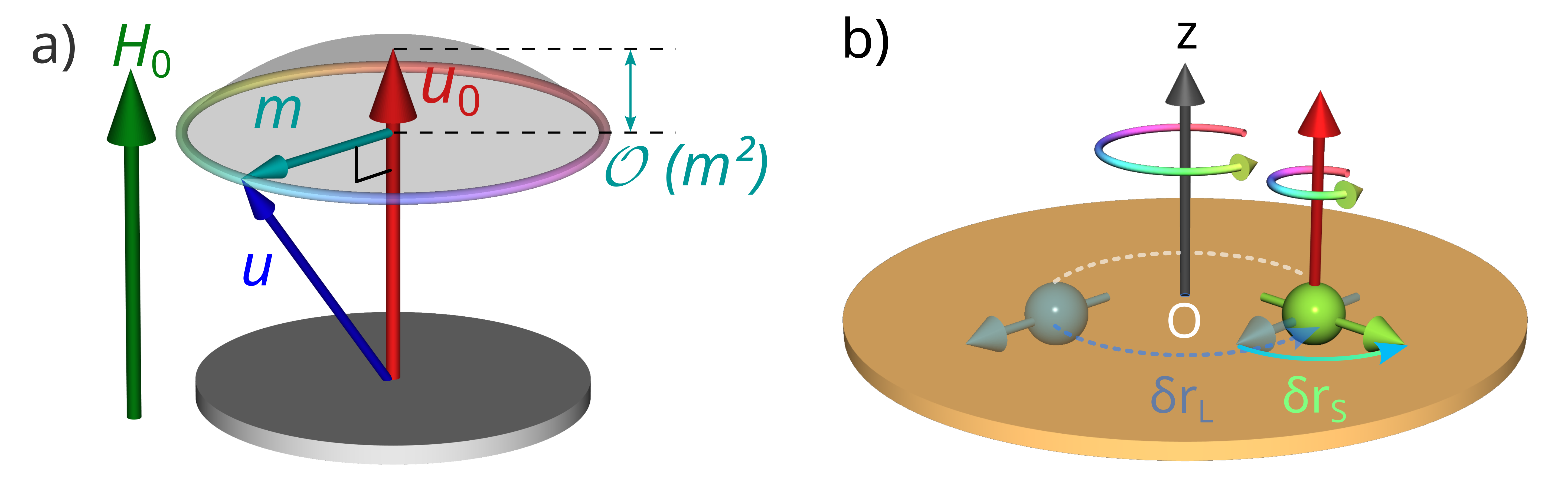}
\caption{\label{fig:geometries} a) Schematic representation of the linear approximation of the Larmor precession. In the out-of-equilibrium regime, the instantaneous magnetization direction $\bm u = {\bm M}/M_{\scriptscriptstyle S}$ (blue vector) rotates counterclockwise around the equilibrium direction $\bm u_0$ (red vector). We define the linear spin-wave $\bm m = \bm u - (\bm u \cdot \bm u_0 ) \bm u_0$, the dynamical deviation component perpendicular to $\bm u_0$ and neglect axial component of $\mathcal{O}(m^2)$.  (a) In axisymmetric geometries, the conservation of total angular momentum follows from the invariance of the action on a global rotation $\delta r_J$ through an infinitesimal angle $\delta \theta $, about a natural $O_z$ axis (black arrow). This rotation admits a decomposition \smash{$\delta r_J = \delta r_L \circ \delta r_S$} as a combination of an \emph{extrinsic} rotation about the origin $O$: the orbital component, and an \emph{intrinsic} rotation in the local frame (red arrow): the spin component.} 
\end{figure}

In this section, we formulate the long wavelength linear spin-wave (SW) dynamics, for a general textured ferromagnet with finite geometry, as a classical field theory. This allows us to define unambiguously the conserved canonical pseudo momenta associated to these SWs when the system exhibits relevant global continuous symmetries, and to discuss the general conditions under which such symmetries may exist. We consider a ferromagnet with homogeneous magnetic properties, far from its Curie temperature, which occupies a region $\Omega$ of volume $V$ with boundary $\Gamma$, surrounded by an empty space region \smash{$\Omega_\infty$} extending to infinity. In accordance with the fundamental assumption of the micromagnetic theory~\cite{brown:1963}, the local instantaneous magnetization vector is defined as \smash{${\bm M}({\bm x},t) = M_{\scriptscriptstyle S} {\bm u}({\bm x}, t)$}, with \smash{$M_{\scriptscriptstyle S}$} the saturation magnetization assumed homogeneous and constant and with \smash{${\bm u}({\bm x}, t)$} a time and space dependent {\em unit} vector field {\em i.e.}, \smash{$||{\bm u}({\bm x}, t)|| = 1$.} Our sole additional hypothesis is that a certain configuration of the magnetization, as specified by a constant unit vector field \smash{${\bm u}_0({\bm x})$}, corresponds to a stationary state {\em i.e.}, to a stable or metastable equilibrium. This equilibrium does not need to be, and in general is not, unique. Considering the small amplitude magnetization dynamics in the vicinity of this equilibrium, we define the dimensionless SW field as the difference between the normalized dynamical magnetization and its equilibrium value {\em i.e.}, \smash{${\bm m} = {\bm u} -  {\bm u}_0$}. It has to be noted that in order to enforce the fundamental assumption of the micromagnetic theory to linear order in $m$, the norm of ${\bm m}$, one has to impose the local and instantaneous constraint \smash{${\bm u}_0 \cdot {\bm m} = 0$}, since we have \smash{$||u|| = 1 - {\bm u}_0 \cdot {\bm m} + O(m^2)$,} as further illustrated in FIG.~\ref{fig:geometries}(a).

\subsection{Lagrangian formulation}
We introduce the following nondimensional Lagrangian density
\begin{equation} \label{eq:lagrangian_tot}
	\mathcal{L} = \frac{1}{2} \frac{\left( {\bm u}_0 \times {\bm m} \right)}{\omega_M} \cdot \partial_t {\bm m} - {\mathcal U} - \kappa \ {\bm u}_0 \cdot {\bm m} , \\
\end{equation}
in which the first term plays the role of a kinetic energy density, with $\omega_M = |\gamma| \mu_0 M_{\scriptscriptstyle S}$, $\gamma < 0$ the gyromagnetic ratio and $\mu_0$ the vacuum permeability. As for ${\mathcal U}$, it includes the dynamic part of the DDI through a contribution \smash{$\mathcal{U}_d =-\frac{1}{2}(\nabla \phi )^2 +\bm{m}\cdot \nabla \phi$}, with the magnetic scalar potential $\phi$ introduced as an auxiliary field; it is implicitly defined by \smash{${\rm U} = \mu_0 M_{\scriptscriptstyle S}^2  \int d^3 x \ {\mathcal U}$} being the second variation of the micromagnetic energy functional at the considered equilibrium~\cite{daquino:2009}, in which the prefactor \smash{$\mu_0 M_{\scriptscriptstyle S}^2$} restores the appropriate physical dimension, and as such being a bilinear functional in terms of ${\bm m}$ and $\phi$. We have to stress that this Lagrangian is {\em local}, thanks to the introduction of $\phi$. Finally, $\kappa$ is a Lagrange multiplier field associated with the previously discussed local orthogonality constraint. We postpone the discussion of dissipation until Sec.~\ref{sec:sym_class}, which is small in most materials of interest, and can be introduced through a suitable Rayleigh dissipation functional if deemed necessary~\cite{gilbert:1955,gilbert:2004}. The action functional is defined as usual by \smash{${\rm S} = \mu_0 M_{\scriptscriptstyle S}^2 \int dt \int d^3 x \ {\mathcal L} = \int dt \ {\rm L}$}, with ${\rm L}$ the Lagrangian. The Euler-Lagrange equations derived from the stationary action principle, with respect to ${\bm m}$, $\phi$ and $\kappa$, yield in respective order
\begin{align}
&\partial_t {\bm m} = \omega_M \ {\bm u}_0 \times \frac{\delta {\rm U}}{\delta {\bm m}}  \label{eq:lin_llg0} \ {\rm in} \ \Omega, \\
&\Delta {\phi} = {\bm \nabla} \cdot {\bm m} \ {\rm in} \ \Omega \cup \Omega_\infty, \label{eq:dyn_poisson} \\
&{\bm u}_0 \cdot {\bm m} = 0 \ {\rm in} \ \Omega, \label{eq:el_ortho}
\end{align}
supplemented by the (natural) Brown's boundary condition \smash{$\partial_{n} {\bm m} = 0$} on $\Gamma$, and the assumption of a vanishing magnetic potential at infinity, insuring well posedness. We recognize these equations, respectively, as the linearized Landau-Lifschitz~(LL) equation in the absence of dissipation~\cite{daquino:2009}, the Poisson equation connecting the magnetic scalar potential to the SW field in the quasi-static limit~\cite{jackson1962,brown:1963}, and the previously mentioned local orthogonality constraint. Hence, we have established that under the stated conditions, the Lagrangian density~(\ref{eq:lagrangian_tot})  defines a classical field theory for the long wavelength linear SW dynamics of a ferromagnet of finite size, in the vicinity of a textured equilibrium ${\bm u}_0$. The consideration of a long wavelength restriction is necessary, since it is clear that the present formulation ceases to be valid for SWs whose wavelength becomes comparable to the dimensions of the underlying magnetic unit cell of the considered crystal. For SWs whose wavelength approaches this microscopic length scale, our starting point {\em i.e.}, the micromagnetic theory, breaks down. It has also to be emphasized that the Lagrangian density~(\ref{eq:lagrangian_tot}) is manifestly invariant under a general coordinate transformation, which is not the case of the well known D\"oring Lagrangian~\cite{doring:1948,miltat:2002,tchernyshyov:2015} for the fully nonlinear micromagnetic dynamics. Finally, it shall be mentioned that our field theory corresponds to a generalization, to finite systems with arbitrary texture, of the Lagrangian formulation previously introduced by Tsukernik~\cite{tsukernik:1966}, in the special case of a uniformly saturated and infinite ferromagnet.

\subsection{Canonical angular momenta} \label{sec:ang_mom}
As we have now formulated the linear dynamics of the SW field as a classical field theory, it becomes in principle possible to apply the Noether's theorem~\cite{noether:1918,weinberg:1995} to derive general expressions for conserved quantities carried by the SWs in presence of invariances of the considered ferromagnet under certain global continuous transformations. This has been discussed already in the nonlinear case~\cite{yan:2013}, with the caveat that the resulting momenta exhibit gauge dependence. This issue can be traced back to the lack of invariance of the considered Lagrangian under a general coordinate transformation~\cite{tchernyshyov:2015}. As previously mentioned, in the linear regime, our Lagrangian density~(\ref{eq:lagrangian_tot}) does not suffer from this limitation. Before applying the Noether's theorem, we first derive it anew succinctly, in a restricted version, putting emphasis on the continuous infinitesimal generators associated to the considered symmetries. In doing so, we obtain a formulation specially well suited to our purpose. For the sake of compactness of notation, we introduce the composite field \smash{${\bm \psi} \equiv \psi^\alpha \equiv (\kappa , \phi, {\bm m})$.} It is assumed that the Lagrangian density only depends on \smash{${\bm \psi}$} and its first order derivatives, with no explicit time dependence {\em i.e.}, \smash{$\mathcal{L} \equiv \mathcal{L}\left( {\bm \psi}, \partial_\mu {\bm \psi} ; {\bm x} \right)$}, with \smash{$\partial_\mu \equiv (\partial_t, \partial_i) \equiv (\partial_t, {\bm \nabla})$.} These conditions are clearly satisfied by the Lagrangian density specified in Eq.~(\ref{eq:lagrangian_tot}), with the explicit space coordinate dependence stemming from the possibility of a nonuniform equilibrium texture. We introduce a general global continuous transformation of the field, {\em in space only} and dependent upon a single real scalar parameter $\xi$, through its associated infinitesimal generator
\begin{equation} \label{eq:inf_gen}
    \delta \tau : {\bm \psi} \rightarrow {\bm \psi} + \frac{\delta {\bm \psi}}{\delta \xi} \delta \xi .
\end{equation}
Expressing that the Lagrangian is invariant upon application of this transformation under the stated hypothesis, we immediately obtain
\begin{equation}
    \frac{\delta {\rm L}}{\delta \xi} = \mu_0 M_{\scriptscriptstyle S}^2 \!\! \int \!\! d^3 x \! \left[ \frac{\partial \mathcal{L}}{\partial {\bm \psi}} \cdot \frac{\delta {\bm \psi}}{\delta \xi} + \frac{\partial \mathcal{L}}{\partial \left(\partial_\mu {\bm \psi} \right)} \cdot \partial_\mu \left( \frac{\delta {\bm \psi}}{\delta \xi} \right) \right] = 0 , \label{eq:nul_var}
\end{equation}
with the Einstein's summation convention being used from now on between repeated indices. We evidently also have
\begin{align}
    \partial_\mu \left(  \frac{\partial \mathcal{L}}{\partial \left(\partial_\mu {\bm \psi} \right)} \cdot \frac{\delta {\bm \psi}}{\delta \xi} \right) &= \nonumber \\\partial_\mu \left(  \frac{\partial \mathcal{L}}{\partial \left(\partial_\mu {\bm \psi} \right)} \right) \cdot \frac{\delta {\bm \psi}}{\delta \xi} +  \frac{\partial \mathcal{L}}{\partial \left(\partial_\mu {\bm \psi} \right)} &\cdot \partial_\mu \left( \frac{\delta {\bm \psi}}{\delta \xi} \right) ,
\end{align}
which allows to rearrange the integrand in Eq.~(\ref{eq:nul_var}) as
\begin{equation}
    \!\!\!\!\!\! \left( \frac{\partial \mathcal{L}}{\partial {\bm \psi}} - \partial_\mu   \frac{\partial \mathcal{L}}{\partial \left( \partial_\mu {\bm \psi} \right)}  \right) \! \cdot \! \frac{\delta {\bm \psi}}{\delta \xi} + \partial_\mu \left(  \frac{\partial \mathcal{L}}{\partial \left(\partial_\mu {\bm \psi} \right)} \! \cdot \! \frac{\delta {\bm \psi}}{\delta \xi} \right) .
\end{equation}
Now, if the time evolution of ${\bm \psi}$ ensues from $\mathcal{L}$, it implies that ${\bm \psi}$ satisfies the associated Euler-Lagrange equations
\begin{equation}
    \frac{\partial \mathcal{L}}{\partial {\bm \psi}} - \partial_\mu   \frac{\partial \mathcal{L}}{\partial \left( \partial_\mu {\bm \psi} \right)} = 0 .
\end{equation}
Hence we derive
\begin{equation} \label{eq:noether0}
     \mu_0 M_{\scriptscriptstyle S}^2 \int d^3 x \ \partial_\mu \left(  \frac{\partial \mathcal{L}}{\partial \left(\partial_\mu {\bm \psi} \right)} \cdot \frac{\delta {\bm \psi}}{\delta \xi} \right)  = 0 ,
\end{equation}
which by application of the divergence theorem, in space only, can be rearranged as
\begin{equation}
\label{eq:noether}
    \frac{d}{dt} Q_{\tau} = - \Phi_\infty = \int_{\Gamma_\infty} \!\! d \sigma ( {\bm n} \cdot {\bm \nabla} \phi) \frac{\delta \phi}{\delta \xi} = 0,
\end{equation}
with the Noether's charge given by
\begin{equation}
    Q_{\tau} =  {\mathcal J}_{\scriptscriptstyle M} \int_\Omega d^3 x \ \left( {\bm u}_0 \times {\bm m} \right) \cdot \frac{\delta {\bm m}}{\delta \xi} , \label{eq:noeth_charge} 
\end{equation} 
in which \smash{$\mathcal{J}_{\scriptscriptstyle M} = M_{\scriptscriptstyle S} / (2 |\gamma| )$} emerges as a natural scale of AM density for the linear SWs. In Eq.~(\ref{eq:noether}), the flux at infinity of the Noether's current {\em i.e.}, \smash{$\Phi_\infty$} is correctly stated to vanish, since the magnetic potential $\phi$ being a solution of Eq.~(\ref{eq:dyn_poisson}) is bound to fall at least as fast as $1/r^2$ when $r \rightarrow \infty$. The equations~(\ref{eq:noether}-\ref{eq:noeth_charge}) constitute a statement of a restricted version of the Noether's theorem establishing the Noether's charge \smash{$Q_{\tau}$} as a globally conserved quantity {\em i.e.}, as a constant of the motion for any on-shell SW field, if the considered textured finite ferromagnetic system exhibits invariance under the continuous field transformation defined by Eq.~(\ref{eq:inf_gen}). It has to be noted that the Noether's theorem is usually derived from a less restrictive hypothesis of the {\em action}, and not the Lagrangian, being invariant under the considered infinitesimal transformation and is stated in a local form as a continuity equation for the Noether's four-current~\cite{soper:2008,weinberg:1995}. However, the canonical Noether's four-current is notoriously not uniquely defined, hence not necessarily physical~\cite{belinfante:1940}. In contrast, the Noether's charge is a uniquely defined and finite quantity for a bounded system, and is properly given by our Eq.~(\ref{eq:noeth_charge}) in the present case. It is noticeable, if not entirely surprising, that this expression of the Noether's charge is completely independent of the considered micromagnetic energy functional, and is obtained as a spatial integral over $\Omega$ only. 

For a finite system, there is no possibility of translational invariance. This leaves the angular momenta, associated with the eventual invariance under some continuous rotations, as the only remaining Noether's charge associated to a spatial symmetry. Let us assume that such invariance is present by continuous rotation around an oriented axis $O_z$, and let \smash{$(O; {\bm e}_r, {\bm e}_\theta, {\bm e}_z )$} be a general cylindrical reference frame.
The canonical AM component projected along $O_z$ is then defined in a standard way as the associated Noether's charge. Applied to the SW field, an infinitesimal rotation \smash{$\delta r_J$} around $O_z$ admits a natural decomposition as \smash{$\delta r_J = \delta r_L \circ \delta r_S = \delta r_S \circ \delta r_L$}, with
\begin{subequations}
\begin{eqnarray}
    \delta r_L &:& {\bm m} \rightarrow {\bm m} - \frac{\partial {\bm m}}{\partial \theta} \delta \theta  ,
    \\
    \delta r_S &:& {\bm m} \rightarrow {\bm m} +  \left( {\bm e}_z \times {\bm m} \right) \delta \theta.
\end{eqnarray}    
\end{subequations}
As illustrated in FIG.~\ref{fig:geometries}(b) for a finite rotation of the field, this corresponds to the natural decomposition of the action of \smash{$\delta r_J$} into an {\em extrinsic} rotation of the point of application of the SW field, under \smash{$\delta r_L$}, in combination with an {\em intrinsic} (or internal) rotation under \smash{$\delta r_S$}. From Eq.~(\ref{eq:noeth_charge}), while considering the Lagrangian density given by Eq.~(\ref{eq:lagrangian_tot}), we immediately obtain the angular momenta along $O_z$, respectively associated to the invariance under \smash{$\delta r_L$} and \smash{$\delta r_S$} as
\begin{align}
    L^z  & =  - \mathcal{J}_{\scriptscriptstyle M} \int_\Omega d^3 x ({\bm u}_0 \times {\bm m}) \cdot \partial_\theta {\bm m } , \label{eq:orbital_momentum} \\
    S^z & =  + \mathcal{J}_{\scriptscriptstyle M} \int_\Omega d^3 x \left( {\bm u}_0 \times {\bm m} \right) \cdot  \left( {\bm e}_z \times {\bm m} \right) , \label{eq:spin_momentum} 
\end{align}  
with the total AM along $O_z$, associated to the invariance under the combined rotation \smash{$\delta r_J$}, obviously given by
\begin{equation}
   J^z = L^z + S^z. \label{eq:total_momentum} 
\end{equation}
It shall be noted that we are considering transformations of the field, while leaving the background material medium {\em i.e.}, the magnetized solid, unaffected. This has for consequence that the derived angular momenta are {\em pseudo} angular momenta in the sense of Refs.~\cite{mcintyre:1981,streib:2021}, which should be carefully distinguished from the \textit{material} angular momenta associated with the rotational invariance of the whole system. The quantity \smash{$S^z$,} respectively \smash{$L^z$,} being associated to the invariance under an internal symmetry, respectively an external one, it is natural to identify it with the SAM of the SW field, respectively its OAM. As a further validation of this identification of \smash{$S^z$}, it is worth considering the elementary case of an infinite and uniformly saturated ferromagnet. In this case, the axis $O_z$ around which the system is invariant under continuous infinitesimal rotation is evidently aligned with \smash{${\bm u}_0$}. Then Eq.~(\ref{eq:spin_momentum}) immediately yields \smash{$S^z = V ( M_{\scriptscriptstyle S} / |\gamma|)(m^2 /2 )$}. Since \smash{$\langle s^z \rangle = - V ( M_{\scriptscriptstyle S} u^z )/ |\gamma|$} is by definition the instantaneous expectation value of the {\em material} SAM along $O_z$ (assuming a negligible contribution of the electronic OAM to the spontaneous magnetization), and since we have $u^z = 1 - m^2/2 + O(m^4)$, we have at leading order in $m$
\begin{equation}
    \langle s^z \rangle \approx \langle s^z \rangle_0 + S^z ,
\end{equation}
where $\langle s^z \rangle _0$ denotes the expectation value of the {\em material} SAM in equilibrium. This is a nontrivial and noticeable identity, since it indicates that the canonical {\em pseudo} SAM carried by the SW field identifies in that case with the loss of {\em material} SAM projected along the symmetry axis, as induced by the SW excitation.

\subsection{Spin-wave eigenmodes}
Since we are considering a linear dynamics, it is suitable to move the problem to the frequency domain, by introducing a complex representation for the real harmonic SW fields and their associated magnetic scalar potential, according to \smash{${\bm m}({\bm x}, t) = \Re [ \tilde{\bm m}({\bm x}) e^{- {\bf i} {\omega} {t}}]$} and  \smash{${\phi}({\bm x}, t) = \Re [ \tilde{\phi}({\bm x}) e^{- {\bf i} {\omega} {t}}]$,} with \smash{${\omega}$} the angular frequency. Inserting this complex representation into Eqs.~(\ref{eq:lin_llg0}-\ref{eq:el_ortho}), which is equivalent to performing a Fourier transform in time, immediately yields  
\begin{subequations}
\begin{align}
    {\bf i} {\omega}_{\scriptscriptstyle M} {\bm u}_0 \times  \frac{\delta {\rm U}}{\delta {\bm m}} [\tilde{\bm m}]&= {\omega} \tilde{\bm m}  \ {\rm in} \ \Omega, \label{eq:spectral_LL} \\
    \Delta \tilde{\phi} - {\bm \nabla} \cdot \tilde{\bm m} &= 0  \ {\rm in} \ \Omega \cup \Omega_\infty, \label{eq:spectral_poisson} \\
    {\bm u}_0 \cdot \tilde{\bm m} &= 0   \ {\rm in} \ \Omega, \label{eq:spectral_ortho}
\end{align}
\end{subequations}
with the previously stated boundary conditions carried over to the complex field amplitudes. Equations~(\ref{eq:spectral_LL}-\ref{eq:spectral_ortho}) define a generalized and constrained boundary value spectral problem. We denote its solutions {\em i.e.}, the SW eigenpairs, as \smash{$\lbrace \omega_\nu, \tilde{\bm m}_\nu\rbrace$}, in which the magnetic scalar potential is being omitted for brevity of notation and since it can also be seen as a dependent quantity implicitly determined by Eq.~(\ref{eq:spectral_poisson}). As previously established~\cite{mills:2006,daquino:2009,naletov:2011}, the eigenpairs satisfy three general properties for a finite ferromagnet with negligeable dissipation: (i) the eigenfrequencies are real and countable, allowing to introduce an countable index $\nu$ as a stand-in for any indexation resulting from the imposition of the boundary conditions, (ii) if $\lbrace \omega_\nu, \tilde{\bm m}_\nu\rbrace$ is a SW eigenpair, then $\lbrace- \omega_\nu, \tilde{\bm m}_\nu^*\rbrace$ is one as well, evidently corresponding to the same physical SW, and (iii) the SW eigenmodes satisfy the orthogonality relation
\begin{equation} \label{eq:sw_ortho}
    \frac {-\bf i}{2 V {\rm sgn} (\omega_\nu)}  \int_{\Omega} d^3 x \left( {\bm u}_0 \times \tilde{\bm m}_\nu^{*} \right) \cdot \tilde{\bm m}_{\nu'}  =   A_\nu^2 \ \delta_{\nu \nu'} , 
\end{equation}
with \smash{$ A_\nu  > 0$} being {\em defined} as a natural norm for the SW eigenmodes. From now on, we will assume the chosen set of SW eigenmodes to be orthonormal in the sense of Eq.~(\ref{eq:sw_ortho}) {\em i.e.}, to satisfy \smash{$A_\nu = 1$.} We will also admit completeness in the relevant function space in which the SW field live, implying that it is possible to uniquely expand any real SW field solution of Eqs.~(\ref{eq:lin_llg0}-\ref{eq:el_ortho}) in terms of the SW eigenmodes, according to
\begin{equation} \label{eq:sw_modexp}
    \!\!\!\!\!{\bm m}({\bm x}, t) = \frac{1}{2} \sum_{\omega_\nu > 0} \left[ b_\nu \tilde{\bm m}_\nu({\bm x}) e^{-{\bf i} \omega_\nu t} + b_\nu^\star \tilde{\bm m}^\star_\nu({\bm x})  e^{{\bf i} \omega_\nu t}\right] , \!
\end{equation}
with the complex modal amplitudes given by
\begin{equation} \label{eq:mod_amp}
    b_\nu = - \frac {\bf i}{V}  \int_{\Omega} d^3 x \left[ {\bm u}_0 \times \tilde{\bm m}_\nu^{*}({\bm x}) \right] \cdot {\bm m} ({\bm x}, 0) .
\end{equation}

\section{Canonical quantization}

In order to perform a canonical quantization of our SW field theory, we need to move from a Lagrangian to a Hamiltonian formulation. The conjugated momentum to the SW field is 
\begin{equation} \label{eq:conj_mom}
	{\bm \pi} = \mu_0 M_{\scriptscriptstyle S}^2 \frac{\partial \mathcal{L}}{\partial (\partial_t {\bm m})} = \mathcal{J}_{\scriptscriptstyle M} {\bm u}_0 \times {\bm m} ,
\end{equation}
from which the classical Hamiltonian is obtained in the standard way, by Legendre transformation of the Lagrangian
\begin{equation}
    {\rm H} = \int \! d^3 x \left( {\bm \pi} \cdot \partial_t \bm{m} - \mu_0 M_{\scriptscriptstyle S}^2 {\mathcal L} \right) .
\end{equation}
As demonstrated in Appendix~\ref{sec:class_ham}, a spectral representation of this classical Hamiltonian can be straightforwardly derived, leading to
\begin{equation} \label{eq:ham_spect}
        {\rm H} =  V {\mathcal J}_{\scriptscriptstyle M} \sum_{\omega_\nu > 0} b_\nu b_\nu^\star \ \omega_\nu .
\end{equation}
In order to proceed further with canonical quantization, one shall recognize that Eq.~(\ref{eq:conj_mom}), together with the orthogonality constraint~(\ref{eq:el_ortho}) defines primary constraints in phase space. Hence, it becomes appropriate to use Dirac's approach for quantization of constrained systems~\cite{dirac:1950,dirac:1964}. As {\em demonstrated} in \mbox{Appendix~\ref{sec:const_quant},} the proper canonical commutation relations at equal times between the different components of the SW field, promoted to field operators in the Heisenberg picture, are then obtained as
\begin{equation} \label{eq:can_comm}
	\left[ \hat{\rm m}^i ({\bm x}, t), \hat{\rm m}^j ({\bm x}', t) \right] = - {\bf i} \ \frac{\hbar}{2 \mathcal{J}_{\scriptscriptstyle M}} \ \varepsilon_{i j k} \ u_0^k \ \delta({\bm x} - {\bm x}') ,
\end{equation}
with \smash{$\varepsilon_{i j k}$} being the totally antisymmetric Levi-Civita symbol. We can then expand the SW field operator in the Schrödinger picture over the eigenmode basis, according to
\begin{equation} \label{eq:can_exp}
	\hat{\bf m}({\bm x}, 0) = \sqrt{\frac{\hbar}{V \mathcal{J}_{\scriptscriptstyle M}}} \sum_{\omega_\nu > 0} \frac{ \tilde{\bm m}_\nu \hat{\rm b}_\nu   + \tilde{\bm m}_\nu^* \hat{\rm b}_\nu^\dag }{2} ,
\end{equation}
which {\em defines} the magnon creation and annihilation operators, \smash{$\hat{\rm b}^\dag_\nu$} and \smash{$\hat{\rm b}_\nu$,} quantum promoted from the complex modal amplitudes according to
\begin{subequations}
\begin{align}
    b_\nu \rightarrow \sqrt{\frac{\hbar}{V \mathcal{J}_{\scriptscriptstyle M}}} \ \hat{\rm b}_\nu ,  \\
    b^\star_\nu \rightarrow \sqrt{\frac{\hbar}{V \mathcal{J}_{\scriptscriptstyle M}}} \ \hat{\rm b}^{\dag}_\nu .
\end{align}
\end{subequations}
The canonical quantization of Eq.~(\ref{eq:mod_amp}) immediately ensues, and it yields 
\begin{subequations}
\begin{align}
    \hat{\rm b}_\nu  &=  -{\bf i} \sqrt{\frac{{\mathcal J}_{\scriptscriptstyle M}}{V \hbar}}  \int_{\Omega} d^3 x \left( {\bm u}_0 \times \tilde{\bm m}_\nu^{*} \right) \cdot \hat{\bf{m}}({\bm x}, 0), \label{eq:mag_cre} \\
    \hat{\rm b}^{\dag}_\nu  &= {\bf i} \sqrt{\frac{{\mathcal J}_{\scriptscriptstyle M}}{V \hbar}}  \int_{\Omega} d^3 x \left( {\bm u}_0 \times \tilde{\bm m}_\nu \right) \cdot \hat{\bf{m}} ({\bm x}, 0). \label{eq:mag_anh}
\end{align}
\end{subequations}
Then, one can straightforwardly {\em prove} that these operators obey the canonical bosonic commutation relations {\em i.e.}, \smash{$  [ \hat{\rm b}_\nu , \hat{\rm b}_{k'} ] = [ \hat{\rm b}_\nu^\dag , \ \hat{\rm b}_{k'}^\dag ] = 0$} and \smash{$[ \hat{\rm b}_\nu , \hat{\rm b}_{k'}^\dag ] = \delta_{k, k'}$}, as a direct consequence of Eq.~(\ref{eq:can_comm}), as detailed in Appendix~\ref{sec:mag_bos}. The canonical quantization of the classical Hamiltonian, in its spectral representation as defined by Eq.~(\ref{eq:ham_spect}), immediately follows by applying the {\em Weyl ordering prescription}~\cite{weyl:1927}
\begin{equation} \label{eq:quant_ham}
    \hat{\rm H} = \frac{1}{2} \sum_{\omega_\nu > 0} \hbar \omega_\nu \left\lbrace \hat{\rm b}^{\dag}_\nu, \hat{\rm b}_\nu \right\rbrace = \sum_{\omega_\nu > 0} \hbar \omega_\nu \left( \hat{\rm N}_\nu + \frac{1}{2} \right) , 
\end{equation}
with \smash{$\lbrace \hat{\rm b}^{\dag}_\nu, \hat{\rm b}_\nu \rbrace$} indicating an anti-commutator and with \smash{$\hat{\rm N}_\nu = \hat{\rm b}^{\dag}_\nu \hat{\rm b}_\nu$} being recognized as the magnon occupation number operator for mode $\nu$. While the zero point energy term in this Hamiltonian {\em i.e.}, the factor $1/2$ inside the parenthesis is sometime removed~\cite{serpico:2024}, in order to avoid dealing with the ultraviolet divergence seemingly occurring due to the SW spectrum being unbounded from above, we believe this is not warranted. As mentioned earlier, our starting point in the form of a linearized continuum micromagnetic theory ceases to be valid for wavelengths $\lambda \lesssim \lambda_c$, with $\lambda_c$ being of the order of the underlying magnetic unit cell size of the constitutive ferromagnetic material. This defines in turn a natural cutoff frequency $\omega_c$ for the SW spectrum, keeping the zero point energy finite. While such an abrupt cutoff strategy is of course a crude approximation, it nevertheless provides a simple prescription which will allow for instance to accurately describe the thermal to quantum transition in the fluctuations of the magnetization field on the basis of the full Hamiltonian~(\ref{eq:quant_ham}), as long as one is only interested in correlations over time scales larger than $1/\omega_c$. This is however a topic beyond the scope of the present paper.  

We want also to outline that previous attempts at the quantization of linear SW theory in finite ferromagnets were either limited to the uniform texture case while {\em postulating} the bosonic commutation relations when promoting the complex mode amplitude to operators~\cite{mills:2006}, or have {\em assumed valid}~\cite{serpico:2024} the fundamental SW field operator commutation relation when considering an arbitrary texture. In the present work, this relation has been demonstrated in the form of our Eq.~(\ref{eq:can_comm}) thanks to Dirac's general approach to constrained systems. Hence, we provide a rigorous and direct quantization of the linearized micromagnetic dynamics, valid for arbitrary finite geometries, equilibrium textures and micromagnetic energy models, with no need to introduce a microscopic localized spin model \`a la Heisenberg.  This puts on a formally solid foundation a general connection between the classical micromagnetic eigenmode theory~\cite{daquino:2009, naletov:2011} and the rapidly growing field of quantum magnonics~\cite{zare:2022,yuan:2022}.

\section{Azimuthal SW modes in axisymmetric ferromagnets}
\label{sec:sym_class}
For the rest of this paper, we will focus our attention on finite ferromagnets sustaining SW with well defined AM. As we have established general expressions for the angular momenta in section~\ref{sec:ang_mom}, we want first to determine the specific conditions under which the total AM is conserved, and what are the necessary additional restrictions for both the OAM and SAM to be separately conserved. In order to do so, we need to consider a specific form for ${\mathcal U}$. Trying to keep it as general as possible, on the basis of the terms most commonly retained in the micromagnetic free energy~\cite{brown:1963,miltat:2002}, we assume that the density of its second variation is given by
\begin{equation} \label{eq:hessian}
    \!\!\!\! {\mathcal{U}=\frac{1}{2} \left[ h_0 m^2 + \lambda _{\rm exc}^2 \nabla\bm{m}:\nabla\bm{m} - h_a (\bm{e}_a \cdot \bm{m}) ^2 \right]  +\mathcal{U}_d}
\end{equation}
in which the first term stems from the equilibrium effective field \smash{$h_0 \bm{u}_0$,} the second term from the exchange interaction with $\lambda _{\rm exc}$ being the exchange length and the third term from a possible uniaxial anisotropy field of strength $h_a$ along the unit vector $\bm{e}_a$.  Evidently, all field quantities have to be understood as having been normalized by \smash{$M_{\scriptscriptstyle S}$.}  

\begin{figure}[ht!]
\includegraphics[width=0.45\textwidth]{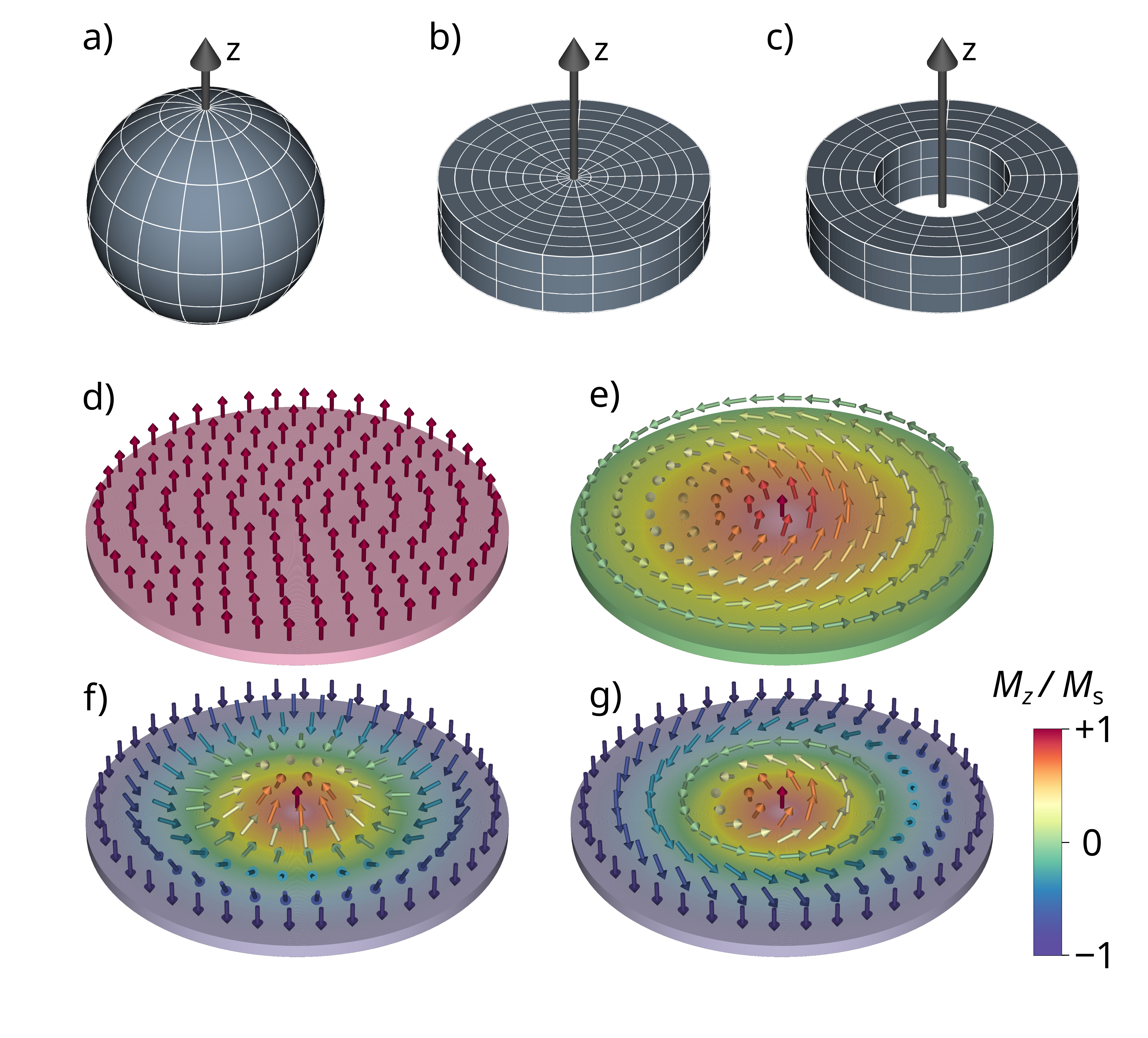}
\caption{\label{fig:geometries2} The notion of conservation of angular momentum is contingent on an invariance under continuous global rotation about a given axis. A first necessary condition for the SW Lagrangian of a finite ferromagnet to satisfy this constraint is, of course, that the volume it occupies in space is axisymmetric. This is for instance the case for ferromagnets in the shape of a) a sphere, b) a disk or c) a ring. The same requirement applies also to its texture. While this is trivially satisfied by d) an axially uniform one, this condition is evidently also fulfilled by {\em e.g.}, e) a vortex and f) a Bloch or g) N\'eel skyrmion. }
\end{figure}

By direct inspection, it is then immediate to determine that the Lagrangian is generically invariant under $\delta r_J$ \mbox{if :} (i) the region $\Omega$, (ii) the equilibrium texture ${\bm u}_0$, and \mbox{(iii) all} the magnetic material properties, are invariant under $\delta r_J$ (see FIG.~\ref{fig:geometries2}). This defines a broad class of {\em axisymmetric ferromagnets}, with different possible shapes including in particular spheres or spheroids~\cite{osada:2016,sharma:2017} as illustrated in FIG.~\ref{fig:geometries2}(a) and  mesoscopic disks or torus as in FIG.~\ref{fig:geometries2}(b-c). It is important to note that the equilibrium texture does not need to be uniform along the axis of symmetry as shown in FIG.~\ref{fig:geometries2}(d), but simply needs to be axisymmetric. This is for instance the case of magnetic microdots in a nonuniform axisymmetric equilibrium, such as a vortex or skyrmion state~\cite{guslienko:2008b,taurel:2016,vukadinovic:2011,rohart:2013}, as sketched in FIG.~\ref{fig:geometries2}(f-g)~\footnote{Even though the interfacial Dzyaloshinskii-Moriya interaction is necessary to stabilize a skyrmion state in a microdot and it is not included in our present model of micromagnetic energy, it can easily be added and it will not break the cylindrical symmetry.}. One can also easily verify that either a non uniformity of the equilibrium texture or a nonnegligeable DDI term~\cite{tsukernik:1966,goldstein:1984} are the two possible origins of a broken invariance under $\delta r_L$ and $\delta r_S$ considered separately under the stated hypothesis. Conversely, if in addition to conditions (i-iii), we also \mbox{have :} (iv) a uniform equilibrium texture aligned with the symmetry axis {\em i.e.}, ${\bm u}_0 = {\bm e}_a = {\bm e}_z$, and (v) a negligible DDI~\footnote{It shall be noted that while a negligible DDI, if combined with condition (iv), is a {\em sufficient} condition for the SW's SAM and OAM to be generically conserved separately, it is {\em not} a necessary condition for this to occur in some particular cases. For instance, some families of Walker's modes~\cite{walker:1957} in saturated spheroids are exact OAM and SAM eigenstates, while the DDI is often a dominant energy term in that case.}, then the Lagrangian is invariant under both $\delta r_L$ and $\delta r_S$. In this case, the SW's OAM and SAM become separately conserved. We propose the name {\em uniaxial exchange ferromagnet} for this more restricted class of axisymmetric ferromagnets.

\begin{figure}[ht!]
\includegraphics[width=0.375\textwidth]{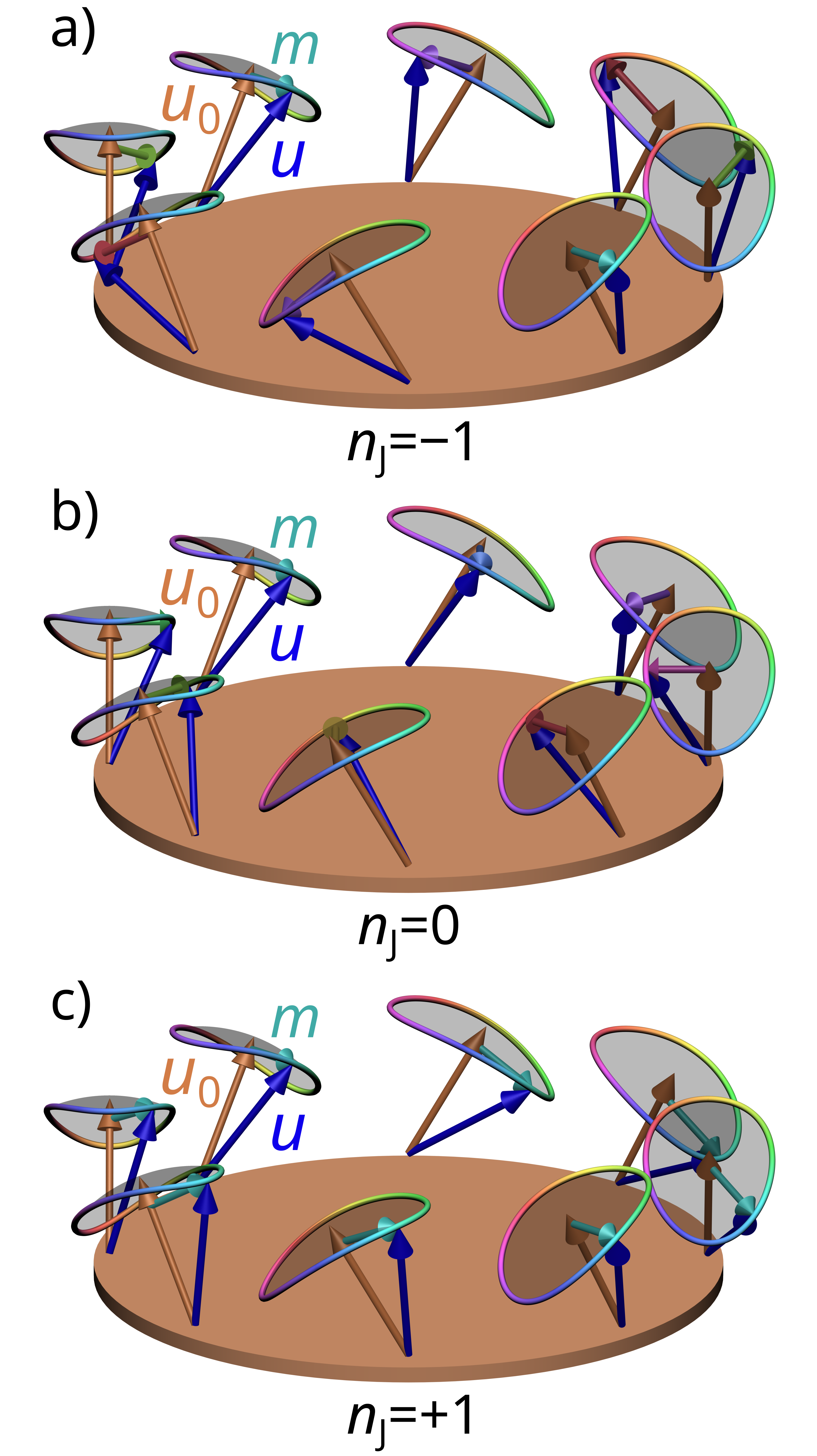}
\caption{\label{fig:azimuthal_SW} Snapshot of the spatio-temporal pattern, $\bm m (x,t)$, formed by azimuthal spin-waves with index $n_{\scriptscriptstyle J}\in [-1,1]$ propagating in a disk axially magnetized with $H_0 < M_{\scriptscriptstyle S}$ (cone state).  The textured local equilibrium is $\bm u_0 (\bm x)$ (orange arrow); the instantaneous magnetization is $\bm u (x,t)$ (blue arrow); the dynamical magnetization vector is $\bm{m} (\bm{x},t)$ (cyan arrow)~\footnote{As illustrated in FIG.~\ref{fig:geometries}, the color wheel encodes the azimuthal direction of $\bm{m}$ when $\bm{u}_0$ is aligned with the normal axis. However, if the magnetization deviates from the normal axis, the color wheel undergoes a local rotation around the radial axis by an angle equal to the polar angle between the magnetization and the normal axis. Consequently, for a given $n_J$, the color pattern remains invariant across any spatial texture.}; the torus indicates its time trajectory along a right-handed rotation around $\bm{u}_0$. By emphasizing the elliptical precession of the magnetization, we point out that the total angular momentum usually cannot be separated into a spin and an orbital component and the sole index $n_{\scriptscriptstyle J}$ is a good quantum number.} 
\end{figure} 

When adopting the model of Eq.~(\ref{eq:hessian}) for the definition of the second variation of the micromagnetic energy functional, Eq.~(\ref{eq:spectral_LL}) becomes
\begin{equation}
    \!\!\!\!\!{\bf i} {\bm u}_0 \times \left[ h_0 \tilde{\bm m} \! - \! \lambda _{\rm exc}^2 \Delta \tilde{\bm m} - h_{\rm a} ({\bm e}_{a} \cdot \tilde{\bm m}) {\bm e}_{a} \! - \!\tilde{\bm h}_d \right] = \omega \tilde{\bm m}\label{eq:spectral_LL2} ,
\end{equation}
with \smash{$\tilde{\bm h}_d = - {\bm \nabla} \tilde{\phi}$} the complex amplitude of the dynamical DDI field. If the system is invariant under $\delta r_J$, it implies that the SW eigenmodes can be chosen so that they are also eigenfunctions of the associated infinitesimal generator, with the following general form 
\begin{subequations}\label{eq:ellipt_sw}
\begin{align}
    \tilde{\bm m}_{\nu, n_{\scriptscriptstyle J}} &=  \breve{\bm m}_{\nu, n_{\scriptscriptstyle J}}(r, z) e^{{\bf i}  n_{\scriptscriptstyle J} \theta } , \\
    {\bm u}_0 \cdot \breve{\bm m} &= 0
\end{align}
\end{subequations}
with the component functions defined in the cylindrical reference frame $\{ \bm{e}_r, \bm{e}_{\theta },\bm{e}_z \}$, \smash{$n_{\scriptscriptstyle J} \in \mathbb{Z}$}, since one can immediately verify that such waves satisfy \smash{$\left[ \left( {\bm e}_z \! \times \cdot \right) - \partial_\theta \right] \tilde{\bm m}_{\nu, n_{\scriptscriptstyle J}} \! = \! { - {\bf i} n_{\scriptscriptstyle J} \tilde{\bm m}_{\nu, n_{\scriptscriptstyle J}}}$}. Here $\nu$ stands for any mode indices, in addition to $n_{\scriptscriptstyle J}$, resulting from the imposition of the Brown's boundary condition~\cite{brown:1963} over the projection of $\Gamma$ on the $(r, z )$ half-plane~\footnote{We do not need to introduce an effective dipolar pinning~\cite{Guslienko2002}}. Hence, in axisymmetric ferromagnets, a natural SW eigenmode basis is constituted by {\em elliptically polarized azimuthal vector waves}, with a dynamical magnetization precessing in the local plane perpendicular to the equilibrium magnetization, as illustrated in FIG.~\ref{fig:azimuthal_SW}. Substituting this waveform into Eq.~(\ref{eq:total_momentum}), while enforcing orthonormality in the sense of Eq.~(\ref{eq:sw_ortho}), immediately yields the total AM of the azimuthal SW modes as
\begin{equation}\label{eq:total_AM_averaged}
J^z_{\nu, n_{\scriptscriptstyle J}} = {\rm sgn}(\omega_{\nu, n_{\scriptscriptstyle J}})  n_{\scriptscriptstyle J} V \mathcal{J}_{\scriptscriptstyle M}  b_{\nu, n_{\scriptscriptstyle J}} b_{\nu, n_{\scriptscriptstyle J}}^\star,
\end{equation} which is found proportional to the azimuthal index $n_{\scriptscriptstyle J}$.  Exactly paralleling the previously performed canonical quantization of the Hamiltonian, the $O_z$ projection of the total AM quantum operator immediately ensues as
\begin{equation}
	\hat{\rm J}^z = \sum_{\omega_{\nu, n_{\scriptscriptstyle J}} > 0} \left( \hat{N}_{\nu, n_{\scriptscriptstyle J}} + \frac{1}{2} \right) n_{\scriptscriptstyle J} \hbar . 
\end{equation}
Hence, the magnon state space can be partitioned into subspaces with distinct AM quantum number $n_{\scriptscriptstyle J}$, in which the total AM expectation value is quantized in units of $n_{\scriptscriptstyle J} \hbar$.

\begin{figure*}[ht!]
\includegraphics[width=0.98\textwidth]{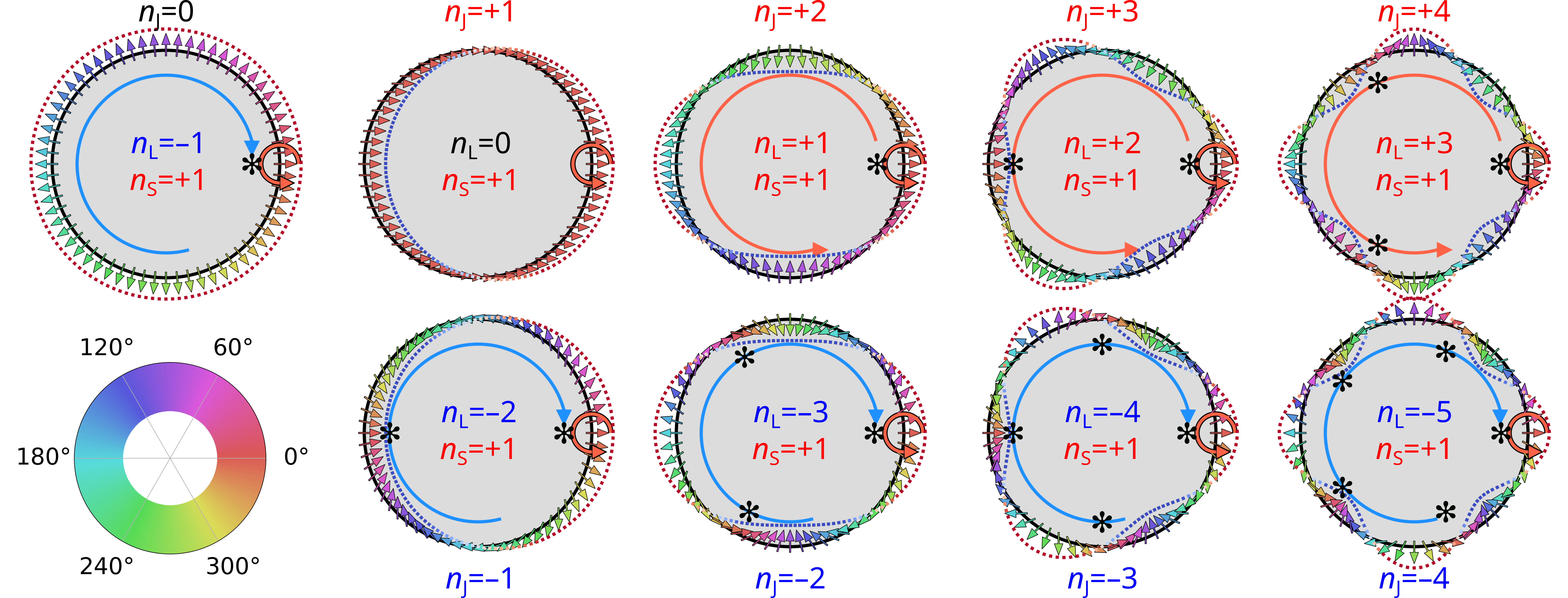}
\caption{\label{fig:normal_SW} (a) Spatial snapshot patterns formed by azimuthal SW modes of index $n_{\scriptscriptstyle J} \in [-4,+4]$ propagating in a magnetic disk uniformly magnetized along the normal direction with the north pole oriented towards the top. In the asymptotic case of $H_0 \gg M_{\scriptscriptstyle S}$, the local precession becomes circular, allowing the decomposition of the total angular momentum into a spin and an orbital component. Amplitude wise, the index $|n_{\scriptscriptstyle J}|$ counts the number of oscillations of the wavefront (visualized here by its radial component; see dotted line) along the periphery, while $|n_{\scriptscriptstyle L}|$ counts the number of revolutions of $\bm{m}$ (see the repetition of the orientation at $0^\circ$, marked by the $\star$ symbol). Polarity wise, each pattern uniquely links the sense of gyration of its wavefront to the direction of the local precession. The large circular arrow indicates the direction of the phase gyration (see color wheel) and the small circular arrow indicates the local Larmor precession.  By convention, a positive index indicates the Larmor direction: right-handed with respect to the magnetization so $n_{\scriptscriptstyle J}=n_{\scriptscriptstyle L}+n_{\scriptscriptstyle S}$. Arranging patterns with the same $|n_{\scriptscriptstyle J}|$ in columns reveals which patterns, having both opposite $n_{\scriptscriptstyle J}$ and opposite frequency within a given column, are coupled by the dynamical DDI.}
\end{figure*}

If the system is additionally invariant under both $\delta r_L$ and $\delta r_S$ considered separately, which practically means that it satisfies all conditions (i-v) listed in section~\ref{sec:sym_class}, the dynamical magnetization lies in the \smash{$(r, \theta)$} plane, perpendicular to the $O_z$ symmetry axis which is aligned with the uniform equilibrium magnetization {\em i.e.}, \smash{${\bm u}_0 = {\bm e}_z$}. The SW eigenmodes can then be chosen as eigenfunctions of both the extrinsic and intrinsic infinitesimal generators of rotation, and their general form is readily obtained as
\begin{equation}
   \tilde{\bm m}_{\scriptscriptstyle \nu, n_{\scriptscriptstyle L}, n_{\scriptscriptstyle S}}\! = \! m_{\scriptscriptstyle \nu, n_{\scriptscriptstyle L}, n_{\scriptscriptstyle S}}(r, z) e^{{\bf i}  n_{\scriptscriptstyle J} \theta } ( {\bm e}_r \! + \! {\bf i} n_{\scriptscriptstyle S} {\bm e}_\theta ), \label{eq:circ_sw}
\end{equation}
with \smash{$n_{\scriptscriptstyle J}  = n_{\scriptscriptstyle L} + n_{\scriptscriptstyle S}$,} \smash{$n_{\scriptscriptstyle S} = \pm 1$} and \smash{$n_{\scriptscriptstyle L} \ \in \ \mathbb{Z}$}, since one can immediately verify the identities \smash{${\bm e}_z \times \tilde{\bm m}_{\scriptscriptstyle \nu, n_{\scriptscriptstyle L}, n_{\scriptscriptstyle S}} = - {\bf i} n_{\scriptscriptstyle S} \tilde{\bm m}_{\scriptscriptstyle \nu, n_{\scriptscriptstyle L}, n_{\scriptscriptstyle S}}$} and \smash{$- \partial_\theta \tilde{\bm m}_{\scriptscriptstyle \nu, n_{\scriptscriptstyle L}, n_{\scriptscriptstyle S}} = - {\bf i} n_{\scriptscriptstyle L} \tilde{\bm m}_{\scriptscriptstyle \nu, n_{\scriptscriptstyle L}, n_{\scriptscriptstyle S}}$}. Hence, in uniaxial exchange ferromagnets, a natural SW eigenmodes basis is constituted by {\em circularly polarized} azimuthal vector waves, with a dynamical magnetization precessing in the plane perpendicular to the symmetry axis, In addition, Eq.~(\ref{eq:spectral_LL2}) reduces in that case into an equation for the scalar amplitude \smash{$m{\scriptscriptstyle \nu, n_{\scriptscriptstyle L}, n_{\scriptscriptstyle S}}(r,z)$}, which can be chosen real without any loss of generality, according to
\begin{equation}
\!\!\!\!\!\! n_{\scriptscriptstyle S} \! \left[ h_0  \! + \!\lambda_\text{exc}^2 \! \left( \frac{n_{\scriptscriptstyle L}^2}{r^2} \! - \!\Delta_{(r,z)} \right) \right] \! m_{\scriptscriptstyle \nu, n_{\scriptscriptstyle L}, n_{\scriptscriptstyle S}} \! = \! \omega_{\scriptscriptstyle \nu, n_{\scriptscriptstyle L}, n_{\scriptscriptstyle S}} m_{\scriptscriptstyle \nu, n_{\scriptscriptstyle L}, n_{\scriptscriptstyle S}} , \!\! \label{eq:ll_spectral0}
\end{equation}
in which \smash{$\Delta_{(r,z)}$} is the Laplacian in cylindrical coordinates, restricted to the $(r,z)$ plane. We can then immediately deduce by direct inspection of Eq.~(\ref{eq:ll_spectral0}) that
\begin{equation}
    \omega_{\scriptscriptstyle \nu, n_{\scriptscriptstyle L}, n_{\scriptscriptstyle S}} = n_{\scriptscriptstyle S} \  \omega_{\scriptscriptstyle \nu, n_{\scriptscriptstyle L}} , \label{eq:rh_sw}
\end{equation}
with \smash{$\omega_{\scriptscriptstyle \nu, n_{\scriptscriptstyle L}} \geq 0$}. This condition stems from the time reversal symmetry breaking associated with the ferromagnetic order. While circularly polarized azimuthal SWs, as defined by Eq.~(\ref{eq:circ_sw}), appear at first sight to exist in both right-handed and left-handed versions, the condition~(\ref{eq:rh_sw}) only allows for azimuthal SWs with right-handed circular polarization in uniaxial exchange ferromagnets. This corresponds of course to the counter-clockwise Larmor precession imposed by our assumption of a negative gyromagnetic ratio, and is further illustrated in FIG.~(\ref{fig:normal_SW}). In addition, it follows from Eq.~(\ref{eq:ll_spectral0}) that the following nontrivial spectral symmetry
\begin{equation}
    \omega_{\scriptscriptstyle \nu, n_{\scriptscriptstyle L}} = \omega_{\scriptscriptstyle \nu, -n_{\scriptscriptstyle L}} 
\end{equation}
shall always be observed in uniaxial exchange ferromagnets. These general properties having been established, we can deduce from Eqs.~(\ref{eq:orbital_momentum}-\ref{eq:spin_momentum}) the following expressions for the OAM and SAM for the SW eigenmodes in this case
\begin{align}
    {L}_{\scriptscriptstyle \nu, n_{\scriptscriptstyle L}, n_{\scriptscriptstyle S}}^z &= n_{\scriptscriptstyle S} n_{\scriptscriptstyle L} V \mathcal{J}_{\scriptscriptstyle M} b_{\scriptscriptstyle \nu, n_{\scriptscriptstyle L}, n_{\scriptscriptstyle S}} b_{\scriptscriptstyle \nu, n_{\scriptscriptstyle L}, n_{\scriptscriptstyle S}}^\star , \\ 
    {S}_{\scriptscriptstyle \nu, n_{\scriptscriptstyle L}, n_{\scriptscriptstyle S}}^z &= V \mathcal{J}_{\scriptscriptstyle M}b_{\scriptscriptstyle \nu, n_{\scriptscriptstyle L}, n_{\scriptscriptstyle S}} b_{\scriptscriptstyle \nu, n_{\scriptscriptstyle L}, n_{\scriptscriptstyle S}}^\star ,  
\end{align}
which are found respectively proportional to \smash{$n_{\scriptscriptstyle S} n_{\scriptscriptstyle L}$} and \smash{$n_{\scriptscriptstyle S}^2 = 1$.} It is immediate to verify that these expressions are compatible with Eq.~(\ref{eq:total_AM_averaged}), as they should. The $O_z$ projections of the OAM and SAM operators verify
\begin{equation}
	\hat{\rm J}^z = \hat{\rm L}^z + \hat{\rm S}^z = \sum_{\nu, n_{\scriptscriptstyle L}} \left( \hat{\rm N}_{\nu, n_{\scriptscriptstyle L}} + \frac{1}{2} \right) (n_{\scriptscriptstyle L} + 1) \hbar . 
\end{equation}
The magnons with orbital quantum number \smash{$n_{\scriptscriptstyle L} = n_{\scriptscriptstyle J} - 1$} carry an OAM $\langle {\rm L}_z \rangle$ quantized in units of $n_{\scriptscriptstyle L} \hbar$, and a SAM $\langle {\rm S}_z \rangle$ quantized in units of $\hbar$.  

Before closing the section, we briefly discuss the influence of dissipation on the conserved quantities. In the presence of Gilbert damping, Eq.~(\ref{eq:lin_llg0}) acquires \smash{$\alpha \bm{u}_0 \times \partial _t \bm{m}$} on the right-hand-side through the Rayleigh dissipation function ${\rm R} = \alpha \mathcal{J}_{\scriptscriptstyle M} \int d^3 x \left| \partial _t \bm{m} \right| ^2 $, where $\alpha >0$ is the Gilbert damping constant. Taking time derivatives of Eqs.~(\ref{eq:orbital_momentum}) and (\ref{eq:spin_momentum}) yields
\begin{align}
    \frac{dL^z}{dt} &= \cdots + 2\alpha \mathcal{J}_{\scriptscriptstyle M} \int _{\Omega }d^3 x \left( \partial _{\theta } \bm{m} \right) \cdot \partial _t \bm{m}  ,  \\
\frac{dS^z}{dt} &= \cdots -2\alpha \mathcal{J}_{\scriptscriptstyle M} \int _{\Omega } d^3 x  \left( \bm{e}_z \times \bm{m} \right) \cdot \partial _t \bm{m}  ,
\end{align}
where $\cdots $ indicate the terms arising from $\rm U$. One can recognize that if $\bm{m}$ is an eigenstate of $\delta r_L$, $\delta r_S$, or $\delta r_J$, the relaxation rate of OAM, SAM, or total AM is given by $2\alpha \omega $ times the respective AM component, generalizing the well-known result for the exchange spin waves.

\section{Semi-analytic SW theory in axially saturated thin microdots}

In this section, we restrict our focus to a specific case of axisymmetric ferromagnet, of special theoretical and experimental relevance {\em i.e.}, magnetic microdots consisting of a magnetic disk of finite thickness $t_\text{disk}$ and radius $R_\text{disk}$. We further limit our discussion to the magnetostatic thin regime in which the aspect ratio $\rho = t_\text{disk} / R_\text{disk} \ll 1$. The axis of symmetry $O_z$ is evidently the disk axis in this case, with the origin of the Cartesian and cylindrical coordinate systems set at the disk center without any loss of generality. We allow for a uniaxial magnetic anisotropy, and a uniform and static external field, both along $O_z$. We introduce the thickness averaged demagnetizing factor along the axial direction {\em i.e.}, $N_z (r)$ with $r$ the radial distance, whose expression can be found in Ref.~\cite{joseph1965}. Then, the additional condition \smash{$h_0 = \left( h^{z} + h_a \right) - N_z(0) > 0$,} insures the existence of an equilibrium magnetization configuration corresponding to a nearly uniform state with \smash{${\bm u}_0 = {\bm e}_z$} almost everywhere except in a narrow region near the disk edge. The associated normalized equilibrium effective field is nearly $z$ independent and is well described by a local approximation~\cite{rohart:2013} which reads
\begin{equation} \label{eq:dot_h0}
h_{0}(r) {\bm e}_z = [ h_0 + \delta N_z(r) ] {\bm e}_z,
\end{equation}
with \smash{$\delta N_z(r) = N_z(0) - N_z(r)$}  defining the inhomogeneous part of the axial equilibrium DDI field, which vanishes on the axis and monotonically increases with $r$, however remaining close to zero almost everywhere in the thin disk limit.

\subsection{Integro-differential formulation}
Having recalled the existence conditions and structure of a uniform equilibrium for the considered disk geometry, we can now formulate in its vicinity the spectral boundary value problem for the linear SW. If the disk thickness is not too large compared to the exchange length, it is clear that the low frequency part of the SW spectrum will be constituted by modes quasi-uniform across the disk thickness. We restrict for now our theory to these modes by integrating Eq.~(\ref{eq:spectral_LL2}) with respect to $z$, while assuming \smash{$\tilde{\bm m}({\bm x}) \equiv \tilde{\bm m}({\bm r}, z) \approx \tilde{\bm m}({\bm r})$}, with ${\bf r}$ the two-dimensional position vector in the disk plane, and the equilibrium effective field to be given by Eq.~(\ref{eq:dot_h0}). We obtain
\begin{equation} \label{eq:freq_llg2} 
    {\bf i} \omega_M {\bf e}_z \times \Big[ ( h_{0}  - \lambda_{\rm exc}^2 \Delta_{\bm r} ) \tilde{\bm m} + ( \delta N_z \tilde{\bm m}
	- {\tilde{\bm h}}_d )  \Big] = \omega \tilde{\bm m} 
\end{equation}
with \smash{$\Delta_{\bm r}$} the two dimensional Laplacian operator in the plane of the disk, while the dynamical DDI field has to be understood as having been averaged over $z$, according to
\begin{equation}
    {\tilde{\bm h}}_d ({\bm r}) = \frac{1}{t_{\rm disk}} \int_{-t_{\rm disk}/2}^{t_{\rm disk}/2} \!\! dz \ \tilde{\bm h}_d ({\bm r}, z) .
\end{equation}
One shall note that from now on the change in arguments is deemed sufficient to distinguish a $z$ averaged quantity from its progenitor. To proceed further, instead of supplementing Eq.~(\ref{eq:freq_llg2}) with Eq.~(\ref{eq:spectral_poisson}) and solving simultaneously for the SW field and magnetic potential, we choose instead to express the dynamical DDI field explicitly in terms of $\tilde{\bm m}$. For that purpose, it is useful to introduce the Green's function for the Laplacian in ${\mathbb R}^3$ {\em i.e.}, ${\rm G}({\bm x}, {\bm x}')$ solution of
\begin{subequations}
\begin{align}
	&\Delta {\rm G}({\bm x}, {\bm x}') = \delta ({\bm x} - {\bm x}') , \label{eq:poisson_gf} \\
	&\!\!\!\!\!\lim_{||{\bm x} - {\bm x}'|| \rightarrow + \infty} {\rm G}({\bm x}, {\bm x}') = 0 , \label{eq:asymptotic_bc_gf}
\end{align}
\end{subequations}
from which one can derive a $z$ averaged in-plane Green's function according to
\begin{equation} \label{eq:perp_gf}
	{\rm G}({\bm r}, {\bm r'}) = \frac{1}{d} \int_{-t_{\rm disk}/2}^{t_{\rm disk}/2} \!\!\! dz \int_{-t_{\rm disk}/2}^{t_{\rm disk}/2} \!\!\! dz' \ {\rm G}({\bm x}, {\bm x}') .
\end{equation}
This allows to formally express the dynamical DDI field as a functional of \smash{$\tilde{\bm m}$,} according to
\begin{equation} \label{eq:integ_hd2}
{\bm h}_d [\tilde{\bm m}]({\bm r}) = {\bm \nabla}_{\bm r} \int_{r' < R_\text{disk}} \!\!\!\! d^2 r' \ \tilde{\bm m}({\bf r}') \cdot {\bm \nabla}_{{\bm r}'} {\rm G}({\bf r}, {\bf r}') ,
\end{equation}
with \smash{${\bm \nabla}_{\bm r}$} the two dimensional gradient operator in the plane of the disk~\footnote{Green's function in $\mathbb{R}^3$ can take only sufficiently regular (and fast-decaying) functions as a source term. Therefore, it is invalid for $\bm{m}\left( \bm{r} \right)$ that varies discontinuously on $\Gamma $ as $\nabla \cdot \bm{m}$ behaves like a delta-function near the boundary, for which a surface term has to be added to $G$~\cite{yamamoto:2023}. However, this surface term is such that it cancels exactly the surface contribution arising from the integration by parts that has been carried out to derive Eq.~(\ref{eq:perp_gf}), insuring the validity of our formulation.}. At this stage it is suitable to complete the adimensionalization of the problem, with the disk radius $R_{\rm disk}$ constituting a natural choice of scaling length. Hence we define a normalized position vector through \smash{$\underline{\bf r} = {\bf r} / R_\text{disk}$,} resulting in the mapping of $\Omega$ onto the unit disk $\mathcal{D}^2$ and of $\Gamma$ onto the unit circle $\mathcal{S}^1$. Consequently and from now on, all differential and integral operators have to be understood to be defined in terms of adimensional coordinates and operating over $\mathcal{D}^2$, and all adimensionalized quantities will be denoted with an underline. We immediately obtain
\begin{equation}
{\bf i} {\bf e}_z \! \times \! \Big\lbrace ( \underline{\omega}_{\scriptscriptstyle K} - \underline{\omega}_{\rm exc} \Delta ) \tilde{\underline{\bm m}} +  \delta \underline{\omega} [\tilde{\underline{{\bm m}}}]  \Big\rbrace = \underline{\omega} \ \tilde{\underline{{\bm m}}} , \label{eq:freq_llg_nd} 
\end{equation}
supplemented by the boundary condition \smash{$\partial_{\underline{r}} \tilde{\underline{{\bm m}}} = 0 \ \ {\rm on} \ \mathcal{S}^1$}, and with
\begin{equation}
    \delta \omega [\tilde{\underline{{\bm m}}}] = \delta \underline{N}_z \tilde{\underline{{\bm m}}} - {\bm \nabla}_{\underline{\bm r}} \!\! \int d^2 \underline{r}' \ \tilde{\underline{{\bm m}}}(\underline{\bm r}') \cdot {\bm \nabla}_{\underline{\bm r}'} \underline{\rm G}(\underline{\bm r}, \underline{\bm r}') ,
\end{equation}
while we have also introduced the adimensional frequency variable \smash{$\underline{\omega} = \omega / \omega_{\scriptscriptstyle M}$}, the adimensional exchange frequency  \smash{$\underline{\omega}_{\rm exc} = \lambda_{\rm exc}^2 / R_\text{disk}^2$} and the adimensional  Kittel's frequency at the disk center \smash{$\underline{\omega}_{\scriptscriptstyle K} = ( {\omega}_{\scriptscriptstyle M} h_0 ) / {\omega}_{\scriptscriptstyle M} = h_0$}. In addition, $\delta \underline{N}_z(\underline{r})$ can be straightforwardly derived from Eq.~(38) of Ref.~\cite{joseph1965} as
\begin{equation} \label{eq:delta_nz}
	\delta \underline{N}_z(\underline{r}) = \int_{0}^{+ \infty} \!\!\!\!\!\! d \xi \  \frac{1 - e^{- \rho \xi}}{\rho \xi} \left[ 1 - J_{0} (\xi \underline{r}) \right] J_{1} (\xi )  ,
\end{equation}
with $J_n$ the Bessel function of the first kind of order $n$, while the $z$ averaged in-plane Green's function in adimensional coordinates reads
\begin{equation} \label{eq:perp_gf_nd}
    \underline{\rm G}(\underline{\bm r}, \underline{\bm r}') = \frac{1}{\rho} \int_{-\frac{\rho}{2}}^{+\frac{\rho}{2}} \!\!\! d\underline{z} \int_{-\frac{\rho}{2}}^{+\frac{\rho}{2}} \!\!\! d\underline{z}' {\rm G}(\underline{\bm x}, \underline{\bm x}') .
\end{equation}
Taken together, Eqs.~(\ref{eq:freq_llg_nd}-\ref{eq:perp_gf_nd}) cast finding the $z$ independent part of the SW eigenmode spectrum of an axially saturated ferromagnetic microdot into an adimensional integro-differential spectral boundary value problem over \smash{$\mathcal{D}^2$.} The universal scaling behavior of this spectral problem is controlled by one geometric parameter, the aspect ratio $\rho$, and by two physical parameters {\em i.e.}, \smash{$\underline{\omega}_{\rm exc}$} and \smash{$\underline{\omega}_{\scriptscriptstyle K}$,} which respectively quantify the strength of the exchange interaction and of the field to magnetization coupling at the microdot center, in comparison with the dynamical DDI.

\subsection{Spectral-Galerkin semi-analytic solution}
To solve the boundary value problem at hand we use the spectral-Galerkin method~\cite{canuto:2006}. To facilitate this treatment, it is convenient to use the language, and introduce notations, appropriate to the context of the spectral theory of linear operators acting on function spaces. We consider the Hilbert space ${\rm L}^2(\mathcal{D}^2)$ of the square integrable functions defined over $\mathcal{D}^2$ and with values in ${\mathbb C}^2 \times \lbrace 0 \rbrace$, equipped with the Hermitian ${\rm L}^2$ inner product. We adopt Dirac's bra-ket notation, with \smash{$\langle {\bm r} | \tilde{\bm m} \rangle \equiv \tilde{\bm m}({\bm r})$.} We further consider a sub-space \linebreak \smash{${\rm H}_B(\mathcal{D}^2) \subset {\rm L}^2(\mathcal{D}^2)$} that we will not further define beyond to say that it contains functions with suitable smoothness and which satisfy Brown's boundary condition {\em i.e.}, homogeneous Neumann boundary condition, on $\mathcal{S}^1$. We admit without further specification or demonstration that it is also a Hilbert space. We can now recast the spectral boundary value problem at hand, initially specified by Eqs.~(\ref{eq:freq_llg_nd}-\ref{eq:perp_gf_nd}), as finding the eigenpairs \smash{$(\underline{\omega}_\nu, |\underline{\tilde{\bm m}}_\nu \rangle)$} in \smash{$\mathbb{R} \times {\rm H}_B(\mathcal{D}^2)$} solutions of 
\begin{equation} \label{eq:full_eigen}
    \hat{\mathcal O} | \underline{\tilde{\bm m}}_\nu \rangle =  \underline{\omega}_\nu | \underline{\tilde{\bm m}}_\nu \rangle ,
\end{equation}
with 
\begin{equation}
    \hat{\mathcal O} = \hat{\mathcal O}_{\rm oe} +  \left( \hat{\mathcal O}_{\rm i}  + \hat{\mathcal O}_{\rm d} \right) ,
\end{equation}
in which
\begin{subequations}
\begin{align}
    \!\!\!\!\!\langle \underline{\bm r} | \hat{\mathcal O}_{\rm oe} | \underline{\tilde{\bm m}} \rangle &= {\bf i} {\bf e}_z \times ( \underline{\omega}_{\scriptscriptstyle K} - \underline{\omega}_{\rm exc} \Delta ) \underline{\tilde{\bm m}} ,\\
    \!\!\!\!\!\langle \underline{\bm r} | \hat{\mathcal O}_{\rm i} | \underline{\tilde{\bm m}} \rangle &= {\bf i} \delta \underline{N}_z {\bf e}_z \times \underline{\tilde{\bm m}} , \label{eq:inho_oper} \\
    \!\!\!\!\!\langle \underline{\bm r} | \hat{\mathcal O}_{\rm d} | \underline{\tilde{\bm m}} \rangle &= - {\bf i} {\bf e}_z \times {\bm \nabla}_{\underline{\bm r}} \!\! \int d^2 \underline{r}' \ \tilde{\underline{{\bm m}}}(\underline{\bm r}') \cdot {\bm \nabla}_{\underline{\bm r}'} \underline{\rm G}(\underline{\bm r}, \underline{\bm r}') , \!\! \label{eq:ddi_oper}
\end{align}
\end{subequations}
correspond in respective order to the linear operators associated with the homogeneous part of the equilibrium effective field together with the dynamical exchange field, the inhomogeneous part of the equilibrium DDI and finally to the dynamical part of the DDI. It is elementary to demonstrate, with the help of the second Green's identity, that \smash{$\hat{\mathcal O}_{\rm oe}$} is self-adjoint. Although \smash{$\hat{\mathcal O}_{\rm oe}$} is not compact, we accept that the eigenvectors of \smash{$\hat{\mathcal O}$}, the exchange-dipole SWs, admit a series expansion in terms of an orthonormal basis of eigenvectors of \smash{$\hat{\mathcal O}_{\rm oe}$}, the exchange only SWs. This particular choice of {\em spectral expansion} is specially appealing, since it already diagonalizes a part of \smash{$\hat{\mathcal O}$.} Additionally, from the results of section~\ref{sec:sym_class}, we recognize that the eigenfunctions of \smash{$\hat{\mathcal O}$} are of the general form specified by Eq.~(\ref{eq:ellipt_sw}), while the eigenfunctions of \smash{$\hat{\mathcal O}_{\rm oe}$} are of the general form specified by Eq.~(\ref{eq:circ_sw}). 

\begin{figure}
\includegraphics[width=0.45\textwidth]{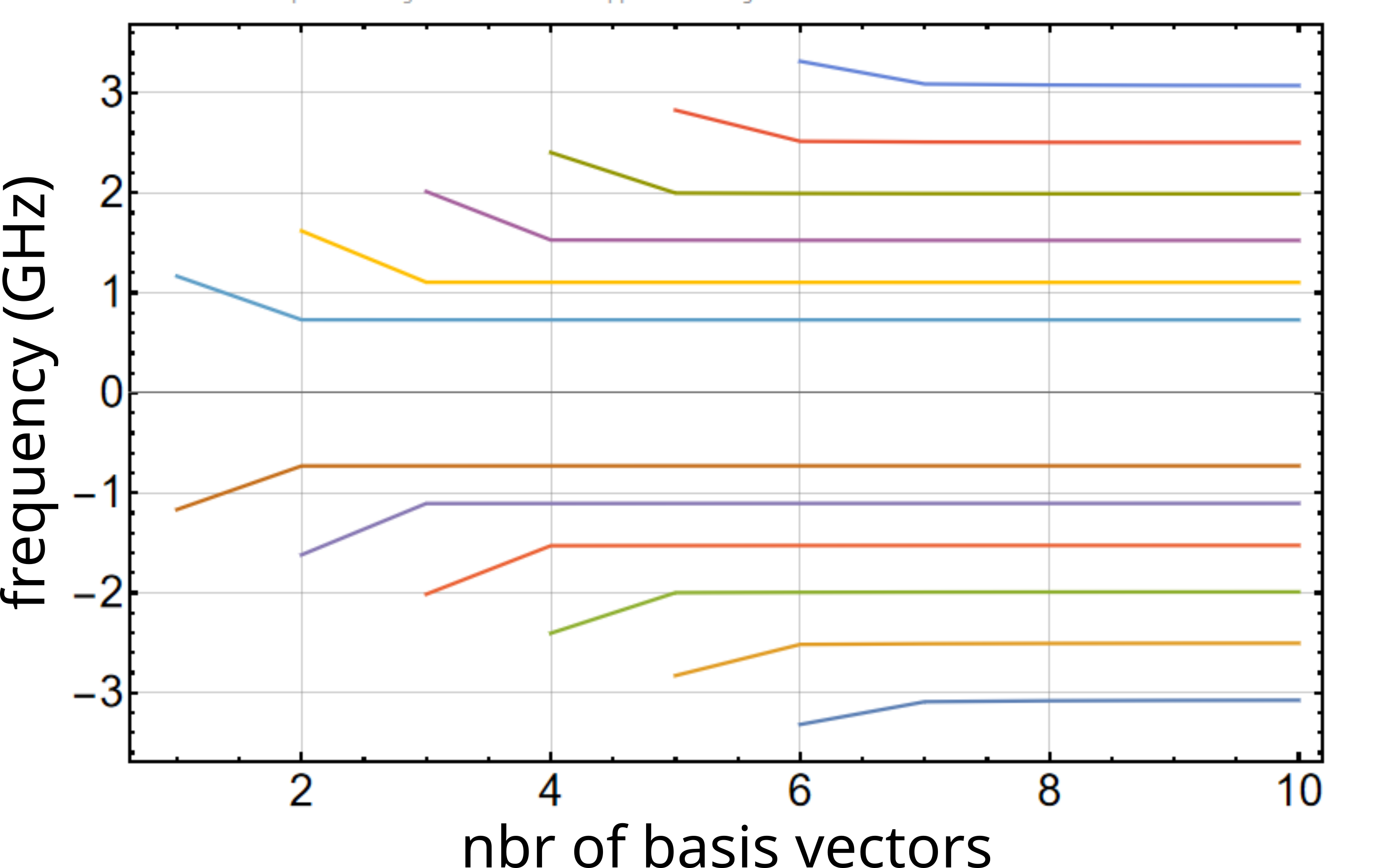}
\caption{\label{fig:7} Convergence in eigenfrequency (expressed in GHz) of the first ten radial harmonics for the exchange-dipole SWs belonging to subspace $n_{\scriptscriptstyle J} = 0$, as a function of the number of basis eigenvectors retained in the truncated expansion.}
\vspace{-10 pt}
\end{figure}

Taking advantage of the orthogonality between eigenspaces labeled with distinct $n_J$ {\em i.e.}, corresponding to SWs carrying distinct quanta of total AM, we write
\begin{equation} \label{eq:hilbert_exp}
    | \underline{\tilde{\bm m}}_{\nu, n_{\scriptscriptstyle J}} \rangle = \sum_{{n}_{\scriptscriptstyle R}} \sum_{n_{\scriptscriptstyle S} = \pm 1} C_{\nu, n_{\scriptscriptstyle J}}^{{n}_{\scriptscriptstyle R}, n_{\scriptscriptstyle S}} \ | \underline{\tilde{\bm m}}_{{n}_{R}, n_{\scriptscriptstyle J} - n_{\scriptscriptstyle S}, n_{\scriptscriptstyle S}}^{\rm (oe)} \rangle ,
\end{equation}
with the amplitude coefficients \smash{$C_{\nu, n_{\scriptscriptstyle J}}^{{n}_{\scriptscriptstyle R}, n_{\scriptscriptstyle S}}$} being complex numbers and with
\begin{equation} \label{eq:eigen=oe}
    \hat{\mathcal O}_{\rm oe} | \underline{\tilde{\bm m}}_{{n}_{R}, n_{\scriptscriptstyle L}, n_{\scriptscriptstyle S}}^{\rm (oe)} \rangle = n_{\scriptscriptstyle S} \underline{\omega}_{{n}_{R}, n_{\scriptscriptstyle L}}^{\rm (oe)} | \underline{\tilde{\bm m}}_{{n}_{R}, n_{\scriptscriptstyle L}, n_{\scriptscriptstyle S}}^{\rm (oe)} \rangle
\end{equation}
in which \smash{$n_{\scriptscriptstyle L} = n_{\scriptscriptstyle J} - n_{\scriptscriptstyle S}$}, and with \smash{$\underline{\omega}_{{n}_{\scriptscriptstyle R}, n_{\scriptscriptstyle L}} >0$} in accordance with Eq.~(\ref{eq:rh_sw}). As detailed in Appendix~\ref{sec:exc_spec}, we have derived analytical expressions for the eigenfrequencies \smash{$\underline{\omega}_{\underaccent{\tilde}{n}_{\scriptscriptstyle R}, n_{\scriptscriptstyle L}}^{\rm (oe)}$} and eigenvectors \smash{$| \underline{\tilde{\bm m}}_{\underaccent{\tilde}{n}_{\scriptscriptstyle R}, n_{\scriptscriptstyle L}, n_{\scriptscriptstyle S}}^{\rm (oe)} \rangle$}. The form of the series expansion given by Eq.~(\ref{eq:hilbert_exp}) allows to consider anew the physical relevance of the negative frequency part of the SW spectrum {\em i.e.}, of the branch $n_{\scriptscriptstyle S}=-1$ for the exchange only SWs. We see now that the inclusion of this negative frequency branch in Eq.~(\ref{eq:hilbert_exp}), within a sub-space with fixed \smash{$n_{\scriptscriptstyle J}$,} is mandatory to enable capturing the general {\em elliptical} local polarization of the full exchange-dipole SWs. It is also appropriate to further comment on the additional mode indices $\nu$ and \smash{${n}_{\scriptscriptstyle R}$.} On one hand, for the exchange only SWs, the index \smash{$n_{\scriptscriptstyle S}$} distinguishes the positive and negative frequency branches of the spectrum, within a sub-space with given \smash{$n_{\scriptscriptstyle J}$.} Hence, the remaining index \smash{$n_{\scriptscriptstyle R}$}, resulting from the imposition of the Brown's boundary condition at the disk edge, can be chosen as a non-negative integer increasing by increasing order of \smash{$\underline{\omega}_{{n}_{\scriptscriptstyle R}, n_{\scriptscriptstyle L} - 1}^{\rm (oe)}$}, to label the different exchange only SW radial harmonics within the sub-space. On the other hand, for the exchange-dipole SWs with given \smash{$n_{\scriptscriptstyle J}$,} there is no available index  distinguishing the positive and negative frequency parts of the spectrum. Hence we index the radial harmonics for the positive frequency part of the spectrum with \smash{$\nu \equiv n_{\scriptscriptstyle R}$,} with  \smash{$n_{\scriptscriptstyle R}$} a non-negative integer increasing by increasing order of \smash{$\underline{\omega}_{{n}_{\scriptscriptstyle R}, n_{\scriptscriptstyle J}}$,} and the negative frequency part of the spectrum with \smash{$\nu \equiv \overline{n_{\scriptscriptstyle R}}$,} with  \smash{$n_{\scriptscriptstyle R}$} a non-negative integer increasing by increasing order of \smash{$|\underline{\omega}_{\overline{{n}_{\scriptscriptstyle R}}, {n_{\scriptscriptstyle J}}}|$.} After substitution of the spectral expansion given by Eq.~(\ref{eq:hilbert_exp}), {\em truncated} to a maximum radial order \smash{${n}_{\scriptscriptstyle R}^{\rm max}$,} into the full problem specified by Eq.~(\ref{eq:full_eigen}), followed by a projection over the same truncated eigenmode basis  \smash{$\forall {n}_{\scriptscriptstyle R} \in \lbrace 0, {n}_{\scriptscriptstyle R}^{\rm max} \rbrace$} and for \smash{$ {n}_{\scriptscriptstyle S} = \pm 1$}, we obtain
\begin{widetext}
    \begin{equation}
    \sum_{n_{\scriptscriptstyle R}' = 0}^{{n}_{\scriptscriptstyle R}^{\rm max}}  \sum_{n_{\scriptscriptstyle S}' = \pm 1} \left[ \left( {O}_{{\rm oe}, n_{\scriptscriptstyle J}}^{n_R, n_{\scriptscriptstyle S}} - \underline{\omega}_{\nu, n_{\scriptscriptstyle J}} \right) \delta_{n_R, n_R'} \delta_{n_{\scriptscriptstyle S}, n_{\scriptscriptstyle S}'}  +  \left( {O}_{{\rm i}, n_{\scriptscriptstyle J}}^{n_R, n_{\scriptscriptstyle S}, n_R', n_{\scriptscriptstyle S}} \delta_{n_{\scriptscriptstyle S}, n_{\scriptscriptstyle S}'} + {O}_{{\rm d}, n_{\scriptscriptstyle J}}^{n_R, n_{\scriptscriptstyle S}, n_R', n_{\scriptscriptstyle S}'} \right) \right] C_{\nu, n_{\scriptscriptstyle J}}^{n_R', n_{\scriptscriptstyle S}'} = 0 ,
\end{equation}
\end{widetext}
or in matrix form
\begin{equation} \label{eq:full_eigen_proj}
    \left( \bar{\bar{O}} - \underline{\omega}_{\nu, n_{\scriptscriptstyle J}} \bar{\bar{1}}  \right) {\bm C}_{\nu, n_{\scriptscriptstyle J}}  = 0,
\end{equation}
with
\begin{equation}
    \bar{\bar{O}} = \bar{\bar{O}}_{\rm oe} +  \left( \bar{\bar{O}}_{\rm i}  + \bar{\bar{O}}_{\rm d} \right)
\end{equation}
in which all matrices involved are square matrices of dimension \smash{$2 ({n}_{\scriptscriptstyle R}^{\rm max} + 1)$}, with $\bar{\bar{1}}$ the identity matrix, and with
\begin{equation}
    {O}_{{\rm oe}, n_{\scriptscriptstyle J}}^{n_R, n_{\scriptscriptstyle S}} = n_{\scriptscriptstyle S} \  \underline{\omega}_{{n}_{R}, n_{\scriptscriptstyle J}-n_{\scriptscriptstyle S}}^{\rm (oe)}
\end{equation}
while the other matrix elements are
\begin{equation} \label{eq:other_matelem}
    \!\!\!\!\! {O}_{{\rm i(d)}, n_{\scriptscriptstyle J}}^{n_R, n_{\scriptscriptstyle S}, n_R', n_{\scriptscriptstyle S}'} \! = \! \langle \tilde{\bm m}_{n_R, n_{\scriptscriptstyle J} - n_{\scriptscriptstyle S}, n_{\scriptscriptstyle S}}^{\rm (oe)} | \hat{\mathcal O}_{i (d)} | \tilde{\bm m}_{n_R', n_{\scriptscriptstyle J} - n_{\scriptscriptstyle S}', n_{\scriptscriptstyle S}'}^{\rm (oe)} \rangle .\!\!\!\!\!
\end{equation}
The derivation of an analytical expression for the matrix elements of \smash{$\hat{\mathcal O}_{i}$}, can be found in Appendix~\ref{sec:ie_matelem}, while it is detailed in Appendix~\ref{sec:ddi_matelem} for the matrix elements of \smash{$\hat{\mathcal O}_{d}$}. The truncation and projection steps complete the spectral-Galerkin mapping of the initial spectral boundary value problem determining the exchange-dipole SWs, as specified by Eq.~(\ref{eq:full_eigen}), into a finite dimensional linear algebra problem {\em i.e.}, the diagonalization of a square matrix of dimension \smash{$2 ({n}_{\scriptscriptstyle R}^{\rm max} + 1)$}, as specified by Eq.~(\ref{eq:full_eigen_proj}). This mapping is being done independently in each sub-space defined by a distinct total AM quantum number  $n_{\scriptscriptstyle J}$. Even though the outlined procedure leads to a {\em dense} matrix diagonalization problem, it can be comfortably handled numerically with standard linear algebra libraries and with modest computational resources even when considering several hundreds basis vectors. For instance, in the rest of this paper we will discuss some results obtained with a numerical implementation based on the Mathematica\texttrademark~\cite{Mathematica} software platform (see supplementary material). For a given choice of \smash{${n}_{\scriptscriptstyle R}^{\rm max}$}, one obtains \smash{$2 ({n}_{\scriptscriptstyle R}^{\rm max} + 1)$} approximate eigenfrequencies and associated eigenvectors belonging to the exchange-dipole SW spectrum of the considered thin disk. One shall note that, since the basis eigenvector are orthonormal in the sense of Eq.~(\ref{eq:sw_ortho}) as outlined in Appendix~\ref{sec:exc_spec}, the same property carries over to the numerically estimated exchange-dipole eigenvectors after a straightforward normalization according to
\begin{equation}
    \underline{\bm C}_{\nu, n_{\scriptscriptstyle J}}
= \frac{{\bm C}_{\nu, n_{\scriptscriptstyle J}}}{\sqrt{{\bm C}_{\nu, n_{\scriptscriptstyle J}}^\star \cdot {\bm C}_{\nu, n_{\scriptscriptstyle J}}}} .
\end{equation} 

It is beyond the scope of this paper to establish rigorously the convergence properties of the outlined spectral-Galerkin procedure, both in terms of the eigenfrencies and eigenvectors. However, we shall stress that this approach is {\em not} akin to a perturbative expansion. Hence, its validity is {\em not} contingent to any smallness criteria for $|| \bar{\bar{O}}_{\rm i}  + \bar{\bar{O}}_{\rm d}||$ in comparison with $|| \bar{\bar{O}}_{\rm oe}||$. Instead, it relies on the expressiveness of spectral expansions such as Eq.~(\ref{eq:hilbert_exp}), which are known to exhibit optimal exponential convergence as long as the target function is ``reasonably'' smooth~\cite{canuto:2006}. 

As a concrete illustration of the excellent convergence observed in practice, we consider a disk with radius $R_\text{disk} = 0.5 \ \mu {\rm m}$ and thickness $t_\text{disk} = 55 \ {\rm nm}$, hence having an aspect ratio $\rho = 0.002$. We choose material parameters typical of the Yttrium Iron garnet (YIG) material {\em i.e.}, $\mu_0 M_{\scriptscriptstyle S} = 0.17 \ {\rm T}$ and $\lambda_{\rm exc} = 15 \ {\rm nm}$, and we assume a negligible uniaxial anisotropy (see Table.\ref{tab:param}). The applied field is $\mu_0 H^z = \mu_0 M_{\scriptscriptstyle S}$. Hence we have in this case $\underline{\omega}_{\scriptscriptstyle K} = 0$ and $\underline{\omega}_{\rm exc} = 9 \times 10^{-4}$. One can see in FIG.~\ref{fig:7} the evolution of the numerically computed exchange-dipole SW eigenfrequencies as a function of \smash{$ {n}_{\scriptscriptstyle R}^{\rm max} + 1$} {\em i.e.}, as a function of the number of basis vectors on each side of the spectrum, in the AM sub-space \smash{$n_J = 0$.} One observes empirically that retaining \smash{${n}_{\scriptscriptstyle R}^{\rm max} + 1$} basis vectors for each value of \smash{$n_S$} is enough to obtain accurate estimates for the first \smash{$({n}_{\scriptscriptstyle R}^{\rm max} -1)$} eigenfrequencies on both sides of the spectrum. 

\begin{figure}
\includegraphics[width=0.49\textwidth]{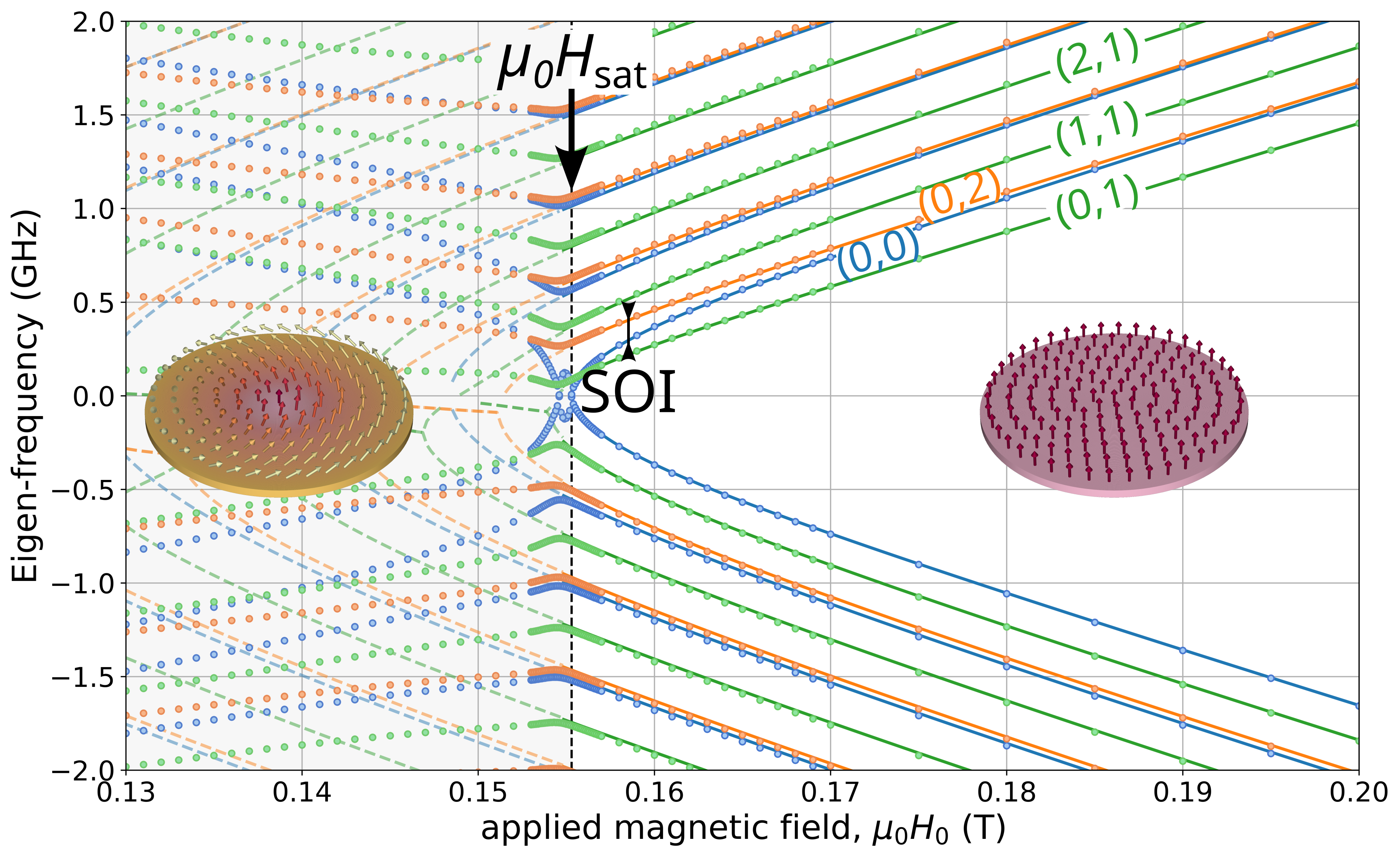}
\caption{\label{fig:8} Evolution as a function of an externally applied magnetic field along the axis of symmetry of exchange-dipole SW radial modes belonging to the subspace $n_{\scriptscriptstyle J} = 0 \ {\rm(in \ blue)}, 1 \ {\rm(in \ green) \ and \ } 2 \ {\rm(in \ orange)}$. The continuous lines are the results of the semi-analytic spectral Galerkin method (see supplementary material) taking full account of the $\hat{\mathcal O}_{d}^{\scriptscriptstyle (-)}$ interaction with higher-order radial modes at fixed $n_{\scriptscriptstyle J}$, while the dots are obtained from finite element numerical simulations for an axially magnetized YIG disk whose parameters are listed in Table \ref{tab:param}. The ensuing SOI is indicated by the black arrow. The validity of the semi-analytical theory ends below the softening of the $(0,0)$ mode at $H_\text{sat}$, which marks the onset of the cone state (see textures).}
\vspace{-10 pt}
\end{figure}

\begin{table}
  \caption{Magnetic parameters of the YIG disk: thickness, radius, exchange length, saturation magnetization, uniaxial anisotropy and gyromagnetic ratio.}
  \begin{ruledtabular}
    \begin{tabular}{c c c c c c}
      $t_\text{disk}$  & $R_\text{disk}$ & $\Lambda_{ex}$ & $\mu_0 M_s$ &  $\mu_0 H_{Ku}$  &  
      $\gamma$  \\
     (nm)  &  (nm) & (nm) &  (T) & (T) & (rad.s$^{-1}$.T$^{-1}$) \\ \hline \\
      55 & 500 & 15 & $0.17$ &  0 & 
      $1.77 \cdot 10^{11}$  \\
    \end{tabular}
  \end{ruledtabular}\label{tab:param}
\end{table}

\subsection{Field evolution of the spectrum and SOI for azimuthal SWs}
As a further validation of the developed spectral-Galerkin method, adopting the same geometrical and material parameters as in the previous sections, we consider now the field evolution of the spectrum, for the three AM sub-spaces \smash{$n_J = 0, 1, 2$}, in the vicinity of $\mu_0 H^z = \mu_0 M_{\scriptscriptstyle S}$~\footnote{The corresponding notebook of the semi-analytical model is provided in the supplementary material.}. As further detailed in an associated paper in relation with the analysis of some experimental data~\cite{valet:2024b}, and as illustrated in FIG.~\ref{fig:8}, it corresponds to a uniquely interesting situation in which the manifestation of a field controlled SW SOI is observed. As the field increases, the coupling between the two sides of the spectrum induced by the dynamical DDI becomes negligible, and the two sub-spaces \smash{$n_J = 0, 2$} asymptotically identifies with the sub-spaces \smash{$n_L = -1, +1$} which become degenerate. Conversely, when the field decreases and approaches from above the critical value \smash{$\mu_0 H_c$} at which there is a loss of stability of the uniform equilibrium texture through the softening of the mode \smash{$n_R = 0, n_J=0$}, the lift in degeneracy between the modes \smash{$n_J = 0, 2$} corresponds to the lift in degeneracy between modes having their OAM respectively anti-parallel and parallel to their SAM, hence to SOI, here induced by the dynamical DDI. As seen on FIG.~\ref{fig:8}, and above the critical field value, there is an excellent agreement between the eigenfrequencies as calculated with the spectral-Galerkin approach and the results obtained with an axisymmetric version of a finite element micromagnetic (FEM) eigensolver which has been itself extensively validated~\cite{naletov:2011,taurel:2016}. The FEM simulations presented in FIG.~\ref{fig:8} are performed on a $R_\text{disk}=500$~nm radius and $t_\text{disk}=55$~nm thickness YIG disk, whose volume has been
meshed adaptively around the core singularity, with the smallest mesh-size being 2.5~nm. The magnetic parameters used in the numerical solver are identical to the ones listed in Table.\ref{tab:param}. A first numerical solver is used to calculate the equilibrium state for different values of the perpendicular magnetic field. The initial state introduced in the simulation is a vortex configuration, with the vortex core pointing towards $+z$. Once the equilibrium is obtained, the eigen-frequencies are calculated by an eigen-solver, which directly diagonalizes the linearized Landau-Lifshitz equation. Of course, discrepancies appear between the two methods at and below the critical field value, since in this regime the equilibrium state becomes a non-uniform cone state. This shows the ability of the spectral-Galerkin approach to provide unique semi-analytical capabilities to analyze and provide insight in the magnonic properties of ferromagnetic microdots, even when dealing with subtle and nontrivial effects such as magnon SOI.


\section{Conclusion}

We have presented a self-contained and general theoretical account of the linear SW dynamics in finite and textured ferromagnetic systems as a classical field theory, with the introduction of a gauge invariant Lagrangian as a generalization of the one previously introduced by Tsukernik~\cite{tsukernik:1966} in the infinite and saturated limit. We have performed a canonical quantization of this classical field theory with constraints following Dirac approach~\cite{dirac:1950}, providing a rigorous generalization of the magnon theory~\cite{holstein:1940} to finite and textured ferromagnets, and offering an alternative and more rigorous path to previous bottom-up approaches~\cite{mills:2006, serpico:2024}. Taking further advantage of our field theoretical approach in the case of axisymmetric systems, with the help of Noether's theorem and our canonical quantization, general expressions of the AM, SAM, and OAM in presence of spatially nonuniform background textures have been obtained, and the connection with azimuthal SW eigenmodes has been elucidated. Their discrete azimuthal indices, introduced as the eigenvalues of infinitesimal generators of the corresponding rotation symmetries, have been proven to be AM quantum number. In the second part of the paper, we have applied our general formalism to the conceptually and experimentally important case of a normally magnetized disk. For a generic model of micromagnetic energy functional in which the exchange interaction, uniaxial anisotropy, external magnetic field, and dipole-dipole interactions are taken into account, a semi-analytical spectral-Galerkin theory of the exchange-dipole SWs has been developed. The accuracy of this approach has been numerically tested against FEM calculations in the vicinity of the loss of stability of the uniform equilibrium state, allowing for an unambiguous identification of magnon SOI as further discussed in a joint paper~\cite{Valet2025} in which recent experimental results, for the low-lying SW eigenmodes in a saturated YIG disk employing a magnetic resonance force microscopy, are analyzed. 

It will be interesting to extend the spectral-Galerkin approach to slightly non-uniform states to explore the softening of certain SW modes towards the transition from the saturated state to a magnetic vortex in the light of AM conservation, as we see that the existing formulation already predicts accurately the critical field at which such a transition occurs. As for our field theoretical formulation of the SW dynamics, one can easily imagine its generalizations to include weak non-linearity to derive a more rigorous description of magnon-magnon interaction processes over complex textured backgrounds in finite samples. We also believe that the study of the transition between the thermal to quantum fluctuation regimes in textured nanomagnets could benefit from proper extensions of our general framework.    

\begin{acknowledgments}
 We sincerely thank Prof. Y. Otani from the University of Tokyo and Prof. C. Serpico from the University Federico II of Naples for their invaluable insights and thought-provoking discussions. This work was partially supported by the EU-project HORIZON-EIC-2021-PATHFINDER OPEN PALANTIRI-101046630; the French Grants ANR-21-CE24-0031 Harmony; the PEPR SPIN - MAGISTRAL ANR-24-EXSP-0004; the French Renatech network; and the REIMEI Research Program of Japan Atomic Energy Agency. K.Y. acknowledges support from JST PRESTO Grant No. JPMHPR20LB, Japan, JSPS KAKENHI (No. 21K13886), and JSPS Bilateral Program Number JPJSBP120245708. We also acknowledge partial support by the Japan Science and Technology Agency (JST) as part of Adopting Sustainable Partnerships for Innovative Research Ecosystem (ASPIRE), Grant Number JPMJAP2410.
\end{acknowledgments}


\appendix

\section{Classical Hamiltonian spectral representation}
\label{sec:class_ham}

In order to derive some equivalent forms of the classical Hamiltonian in a precise sense, it is necessary to distinguish weak equalities from strong ones, even though we are not making this distinction for the rest of this work. Within the scope of this section, we will say that two functionals of the fields, say ${\rm F}$ and ${\rm G}$,  are weakly equal if they are equal when the fields are solutions of the Euler-Lagrange equations under all relevant constraints, but generally not equal for arbitrary realizations of the fields. This will be denoted \smash{${\rm F} \approx {\rm G}$.} If ${\rm F}$ and ${\rm G}$ are equal for arbitrary realizations of the fields, they will be stated to be strongly equal, and we will write \smash{${\rm F} = {\rm G}$.} 

From the conjugated momentum to the SW field, as given by Eq.~(\ref{eq:conj_mom}), the Hamiltonian ${\rm H}$ is obtained in the standard way by Legendre transformation of the Lagrangian. This yields
\begin{equation}
    {\rm H} = \int \! d^3 x \left( {\bm \pi} \cdot \partial_t m - \mu_0 M_{\scriptscriptstyle S}^2 {\mathcal L} \right) \approx {\rm U} ,
\end{equation}
with \smash{${\rm U} = \mu_0 M_{\scriptscriptstyle S}^2 \int \! d^3 x \  {\mathcal U}$.} Let us focus now on the first term of the Lagrangian density~(\ref{eq:lagrangian_tot}), which plays the role of a kinetic energy density. Performing a volume integration, we obtain
\begin{subequations}
\begin{align}
    {\rm T} &= \mu_0 M_{\scriptscriptstyle S}^2 \int \! d^3 x \ \frac{1}{2} \frac{\left( {\bm u}_0 \times {\bm m} \right)}{\omega_M} \cdot \partial_t {\bm m} , \label{eq:kin_en}\\
    &\approx \frac{\mu_0 M_{\scriptscriptstyle S}^2}{2}  \int \! d^3 x \ {\bm m} \cdot \frac{\delta {\mathcal U}}{\delta {\bm m}} = {\rm U} ,
\end{align}
\end{subequations}
in which the second (weak) equality has been obtained by assuming the time derivative of the SW field to be given by the right-hand-side of the LL equation~(\ref{eq:lin_llg0}), and by enforcing the orthogonality constraint~(\ref{eq:el_ortho}), while the final (strong) equality results from ${\mathcal U}$ being the density of the second variation of the micromagnetic energy functional. Hence we have established in all generality {\em i.e.}, independently of the specific geometry and of the particulars of the micromagnetic energy functional, that the linear SW Hamiltonian satisfy the remarkable identities
\begin{equation}
    {\rm H} \approx {\rm U} \approx {\rm T} ,
\end{equation}
which relates to the Lagrangian being {\em first order} in time, hence lacking ``inertia''. Having established the weak equality between the classical Hamiltonian and its ``kinetic'' energy part, it is now suitable to substitute in its expression, as given by Eq.~(\ref{eq:kin_en}), the expansion of the SW field in terms of its eigenmodes as given by Eq.~(\ref{eq:sw_modexp})
while taking into account the mode orthogonality relation given by Eq.~(\ref{eq:sw_ortho}) ; then, one immediately obtains
\begin{equation} \label{eq:ham_mod}
    {\rm H} \approx V {\mathcal J}_{\scriptscriptstyle M} \sum_{\omega_\nu > 0} b_\nu b_\nu^\star \ \omega_\nu ,
\end{equation}
which completes the rigorous derivation of Eq.~(\ref{eq:ham_spect}).

\section{Canonical quantum commutators under constraints} 
\label{sec:const_quant}

From Eqs.~(\ref{eq:el_ortho}) and (\ref{eq:conj_mom}), we define the associated primary constraint functions, respectively
\begin{subequations}
\begin{eqnarray}
    \varphi^0 &=& u_0^k m_k , \\
    \varphi^i &=& \pi^i - \mathcal{J}_{\scriptscriptstyle M} \varepsilon_{i k l} u_0^k m^l ,
\end{eqnarray}
\end{subequations}
with \smash{$\varepsilon_{i k l}$} the totally antisymmetric Levi-Civita symbol. It is immediate to verify that these are second class constraints, since their Poisson's brackets at equal times read
\begin{equation}
    \left\lbrace \varphi^a , \varphi^b \right\rbrace_{\scriptscriptstyle PB} = - \left\lbrace \varphi^b , \varphi^a \right\rbrace_{\scriptscriptstyle PB} = C^{a b} \delta ({\bm x} - {\bm x}') ,
\end{equation}
with the components of the matrix \smash{$\bar{\bar C}$} given by
\begin{subequations}
\begin{eqnarray}
    C^{a a} &=& 0 , \\
    C^{0 i} &=& -C^{i 0} = u_0^i , \\
    C^{i j} &=& 2 \mathcal{J}_{\scriptscriptstyle M} \varepsilon_{i j k} u_0^k .
\end{eqnarray}
\end{subequations}
The inverse of this matrix is then readily obtained, and its components read
\begin{subequations}
\begin{eqnarray}
    (\bar{\bar C}^{-1})^{a a} &=& 0 , \\
    (\bar{\bar C}^{-1})^{0 i} &=& - (\bar{\bar C}^{-1})^{i 0} = - u_0^i , \\
    (\bar{\bar C}^{-1})^{i j} &=& - \frac{\varepsilon_{i j k} u_0^k}{2 \mathcal{J}_{\scriptscriptstyle M}}  .
\end{eqnarray}
\end{subequations}
We can then compute the Dirac bracket between different components of the SW field, in accordance with its standard definition {\em i.e.},
\begin{align}
    \!\!\!\!\left\lbrace m^i , m^j \right\rbrace_{\! \scriptscriptstyle DB} \!\!&= \!\!\left\lbrace m^i , m^j \right\rbrace_{\! \scriptscriptstyle PB} \!\! - \! \left\lbrace m^i , {\bm \varphi} \right\rbrace_{\! \scriptscriptstyle PB}^{T} \!\! \bar{\bar C}^{-1} \!\! \left\lbrace {\bm \varphi}, m^j  \right\rbrace_{\! \scriptscriptstyle PB} \nonumber \\
    &= - \frac{\varepsilon_{i j k} u_0^k}{2 \mathcal{J}_{\scriptscriptstyle M}} \delta ({\bm x} - {\bm x}').
\end{align}
Therefore, the correct implementation of canonical quantization under constraints dictates the following quantum commutation relations between the different components of the SW field, promoted to field operators in the Heisenberg picture and taken at equal times
\begin{align} \label{eq:loc_com}
    \left[\hat{m}^i({\bm x}, t), \hat{m}^j({\bm x}', t) \right] &= {\bm i} \hbar \left\lbrace m^i , m^j \right\rbrace_{\! \scriptscriptstyle DB} \nonumber \\
    &= - \frac{{\bm i} \hbar}{2 \mathcal{J}_{\scriptscriptstyle M}} \varepsilon_{i j k} u_0^k \delta ({\bm x} - {\bm x}') .
\end{align}
This completes the rigorous derivation of Eq.~(\ref{eq:can_comm}), and establishes any pair of mutually orthogonal components of the quantum SW field operator, in the plane locally perpendicular to the equilibrium texture, as conjugate quantum variables in real space. One could alternatively express Eq.~(\ref{eq:loc_com}) as
\begin{equation}
    \varepsilon_{i j k} \left[\hat{m}^i({\bm x}, t), \hat{m}^j({\bm x}', t) \right] u_0^k({\bm x}) = - \frac{{\bm i} \hbar}{ \mathcal{J}_{\scriptscriptstyle M}} \delta ({\bm x} - {\bm x}') ,
\end{equation}
making manifest its invariance under general changes of coordinates.

\section{Commutation relations of the magnon operators} \label{sec:mag_bos}
From the definitions of the magnon creation and annihilation operators provided by Eqs.~(\ref{eq:mag_cre}-\ref{eq:mag_anh}), we obtain immediately
\begin{widetext}
\begin{align}
    \left[ \hat{\rm b}_\nu , \hat{\rm b}^{\dag}_{\nu'} \right] &=  \frac{{\mathcal J}_{\scriptscriptstyle M}}{V \hbar} \int_{\Omega} d^3 x \int_{\Omega} d^3 x' \varepsilon_{i k l} u_0^k [\tilde{m}^l_\nu({\bm x})]^\star \varepsilon_{j o p} u_0^o \tilde{m}^p_{\nu'}({\bm x}') \left[ \hat{\rm m}^i({\bm x}, 0), \hat{\rm m}^j({\bm x}', 0)\right] \nonumber \\
    &= -\frac{i}{2 V} \int_{\Omega} d^3 x \varepsilon_{i k l} u_0^k [\tilde{m}^l_\nu({\bm x})]^\star \varepsilon_{j o p} \varepsilon^{i j q} u_0^o \tilde{m}^p_{\nu'}({\bm x}) ({\bm u}_0)_q \nonumber \\
    &=  -\frac{i}{2 V} \int_{\Omega} d^3 x \varepsilon_{i k l} u_0^k [\tilde{m}^l_\nu({\bm x})]^\star (\delta_o{}^q \delta_p{}^i - \delta_o{}^i \delta_p{}^q) u_0^o \tilde{m}^p_{\nu'}({\bm x}) ({\bm u}_0)_q \\
    &=  -\frac{i}{2 V} \int_{\Omega} d^3 x \left( {\bm u}_0 \times \tilde{\bm m}_\nu^{*} \right) \cdot \tilde{\bm m}_{\nu'} + \cancel{ \frac{i}{2 V} \int_{\Omega} d^3 x \left[{\bm u}_0\cdot \left( {\bm u}_0 \times \tilde{\bm m}_\nu^{*} \right) \right] \left( {\bm u}_0 \cdot \tilde{\bm m}_{\nu'} \right) } \nonumber \\
    &= \delta_{\nu \nu'} \nonumber ,
\end{align}
\end{widetext}
in which the second line results from the canonical commutation relation~(\ref{eq:can_comm}), the third line from the general properties of the Levi-Civita symbol, and the last line from the orthonormality of the eigenmodes in the sense of Eq.~(\ref{eq:sw_ortho}). The other two canonical bosonic commutation relations for the magnon creation and annihilation operators can be straightforwardly proven following the same steps.

\section{Eigenfrequencies and eigenvectors for the exchange SWs in the case of an axially saturated thin microdot}
\label{sec:exc_spec} 

\begin{table*}[t]
    \vspace{30pt}
    \centering
    \scalebox{1.15}{
    \begin{tabular}{c|c|c|c|c}
    \toprule
     \diagbox{$|n_{\scriptscriptstyle L}|$}{$n_R$} \ & 0 & 1 & 2 & 3  \\ 
     \midrule
     0 & 0. & 3.8317059702075125 & 7.015586669815632 & 10.173468135062722  \\
     1 & 1.841183781340659 & 5.331442773525031 & 8.536316366346288 & 11.706004902592063  \\
     2 & 3.0542369282271404 & 6.706133194158461 & 9.969467823087447 & 13.170370856016122  \\      
     3 & 4.201188941210528 & 8.01523659837595 & 11.345924310742971 & 14.585848286167023  \\            
	 4 & 5.317553126083994 & 9.28239628524162 & 12.68190844263889 & 15.96410703773155  \\
	 5 & 6.415616375700238 & 10.519860873772254 & 13.9871886301403 & 17.312842487884627  \\
     \bottomrule
    \end{tabular}}
    \caption{The non-trivial positive roots $\alpha_{n_R, n_{\scriptscriptstyle L}}$ of $J'_{n_{\scriptscriptstyle L}}$, for $|n_{\scriptscriptstyle L}| \leq 5$ and $n_R \leq 3$.  \label{tab:roots_lk}}
    \vspace{0pt}
\end{table*}

We are looking for the eigenfrequencies and eigenvectors solution of Eq.~(\ref{eq:eigen=oe}), the latter being of the general form specified by Eq.~(\ref{eq:circ_sw}). Expressed in differential form, this translates into a purely radial boundary value spectral problem over \smash{$[ 0, 1]$}
\begin{equation}
    \!\!\!\!\left[ \frac{d^2}{d \underline{r}^2} + \frac{1}{\underline{r}} \frac{d}{d \underline{r}} + \left( \frac{\underline{\omega}^{\rm (oe)}_{n_R, n_{\scriptscriptstyle L}} - \underline{\omega}_{K}}{\underline{\omega}_{\rm exc}} -  \frac{n_{\scriptscriptstyle L}^2}{\underline{r}^2} \right) \right] m^{\rm (oe)}_{n_R, n_{\scriptscriptstyle L}} = 0 ,  \label{eq:freq_llg_nd31} 
\end{equation}
for the {\em real} eigenfunctions \smash{$m^{\rm (oe)}_{n_R, n_{\scriptscriptstyle L}}$} and the {\em real and positive} eigenfrequencies \smash{$\underline{\omega}^{\rm (oe)}_{n_R, n_{\scriptscriptstyle L}}$,} with the Brown's boundary condition at the disk edge imposing \smash{$ d  m^{\rm (oe)}_{n_R, n_{\scriptscriptstyle L}} / d \underline{r} = 0$} at \smash{$\underline{r} = 1$}. If \smash{$\underline{\omega}^{\rm (oe)}_{n_R, n_{\scriptscriptstyle L}} > \underline{\omega}_K$}, we can introduce the real coordinate transform \smash{$\underline{r}_{\scriptscriptstyle (+)} = \alpha_{\scriptscriptstyle (+)} \underline{r}$} with
\begin{equation}
    \alpha_{\scriptscriptstyle (+)} = \sqrt{\frac{\underline{\omega}^{\rm (oe)}_{n_R, n_{\scriptscriptstyle L}} - \underline{\omega}_K}{\underline{\omega}_{\rm exc}}} , \label{eq:trans+}
\end{equation}
which maps Eqs.~(\ref{eq:freq_llg_nd31}) into
\begin{equation}
    \!\!\!\!\left[ \frac{d^2}{d \underline{r}_{\scriptscriptstyle (+)}^2} + \frac{1}{\underline{r}_{\scriptscriptstyle (+)}} \frac{d}{d \underline{r}_{\scriptscriptstyle (+)}} + \left( 1 - \frac{n_{\scriptscriptstyle L}^2}{\underline{r}_{\scriptscriptstyle (+)}^2} \right) \right] m^{\rm (oe)}_{n_R, n_{\scriptscriptstyle L}} = 0 , \label{eq:freq_llg_nd41} 
\end{equation}
over \smash{$[ 0, \alpha_{\scriptscriptstyle (+)}]$}, with \smash{$ d  m^{\rm (oe)}_{n_R, n_{\scriptscriptstyle L}} / d \underline{r}_{\scriptscriptstyle (+)} = 0$} at \smash{$\underline{r}_{\scriptscriptstyle (+)} = \alpha_{\scriptscriptstyle (+)}$}. Similarly, if \smash{$\underline{\omega}^{\rm (oe)}_{n_R, n_{\scriptscriptstyle L}} < \underline{\omega}_K$}, we can introduce the real coordinate transform \smash{$\underline{r}_{\scriptscriptstyle (-)} = \alpha_{\scriptscriptstyle (-)} \underline{r}$} with
\begin{equation}
    \alpha_{\scriptscriptstyle (-)} = \sqrt{\frac{\underline{\omega}_K - \underline{\omega}^{\rm (oe)}_{n_R, n_{\scriptscriptstyle L}}}{\underline{\omega}_{\rm exc}}} , \label{eq:trans-}
\end{equation}
which maps Eqs.~(\ref{eq:freq_llg_nd31}) into
\begin{equation}
    \!\!\!\!\left[ \frac{d^2}{d \underline{r}_{\scriptscriptstyle (-)}^2} + \frac{1}{\underline{r}_{\scriptscriptstyle (-)}} \frac{d}{d \underline{r}_{\scriptscriptstyle (-)}} - \left( 1 + \frac{n_{\scriptscriptstyle L}^2}{\underline{r}_{\scriptscriptstyle (-)}^2} \right) \right] m^{\rm (oe)}_{n_R, n_{\scriptscriptstyle L}} = 0 , \label{eq:freq_llg_nd51} 
\end{equation}
over \smash{$[ 0, \alpha_{\scriptscriptstyle (-)}]$}, with \smash{$ d  m^{\rm (oe)}_{n_R, n_{\scriptscriptstyle L}} / d \underline{r}_{\scriptscriptstyle (-)} = 0$} at \smash{$\underline{r}_{\scriptscriptstyle (-)} = \alpha_{\scriptscriptstyle (-)}$}. We immediately recognize Eq.~(\ref{eq:freq_llg_nd41}) as the Bessel's equation, and Eq.~(\ref{eq:freq_llg_nd51}) as the modified Bessel's equation. However, all modified Bessel's functions with a finite amplitude at the origin {\em i.e.}, the modified Bessel's functions of the first kind, are strictly growing function over ${\mathbb R}_{+}$, hence cannot satisfy the boundary condition at \smash{$\underline{r}_{\scriptscriptstyle (-)} = \alpha_{\scriptscriptstyle (-)}$}. This establishes that the Kittel's frequency defines a lower bound for the possible eigenfrequencies after having restored physical dimensions, with the exchange SWs spectrum within the stated hypothesis being entirely determined by Eqs.~(\ref{eq:freq_llg_nd41}). Hence in combination with Eq.~(\ref{eq:circ_sw}), we immediately obtain  the general expression of the eigenfunctions \smash{$\forall  \ n_{\scriptscriptstyle L} \in {\mathbb Z}$} with \smash{$n_{\scriptscriptstyle S} = \pm 1$} and \smash{$n_{\scriptscriptstyle J} = n_{\scriptscriptstyle L} + n_{\scriptscriptstyle S}$}, as
\begin{equation} 
   \!\!\underline{\tilde{\bm m}}_{{n}_{R}, n_{\scriptscriptstyle L}, n_{\scriptscriptstyle S}}^{\rm (oe)} \!=\! \frac{A_{n_R, n_{\scriptscriptstyle L}}}{2 \sqrt{\pi}} \! \left( {\bf e}_{r}  + {\bf i} n_{\scriptscriptstyle S} {\bf e}_\theta \right) e^{{\bf i} n_{\scriptscriptstyle J} \theta} J_{n_{\scriptscriptstyle L}} \!\left( \alpha_{n_R, n_{\scriptscriptstyle L}} \underline{r} \right) , \label{eq:exc_SW21}
\end{equation}
in which $J_{n_{\scriptscriptstyle L}}$ is the Bessel's function of the first kind of order $n_{\scriptscriptstyle L}$ and $A_{n_R, n_{\scriptscriptstyle L}}$ is a normalization constant. While, from Eq.~(\ref{eq:trans+}), the eigenfrequencies are given by
\begin{equation}
    \underline{\omega}^{\rm (oe)}_{n_R, n_{\scriptscriptstyle L}} =  1 + \underline{\omega}_{\rm exc} \ \alpha_{n_R, n_{\scriptscriptstyle L}}^2 , \label{eq:exc_SW22} 
\end{equation}
the admissible values of $\alpha_{n_R, n_{\scriptscriptstyle L}}$ are the {\em nontrivial} roots $\alpha_{n_R, n_{\scriptscriptstyle L}}$ of the secular equation deriving from the Brown's boundary condition at the disk edge {\em i.e.},
\begin{equation} \label{eq:exch_sec}
    J'_{n_{\scriptscriptstyle L}} \left( \alpha_{n_R, n_{\scriptscriptstyle L}} \right) = 0 , 
\end{equation}
in which $n_R$ can be chosen so that it indexes the roots in order of increasing value for a given $n_{\scriptscriptstyle L}$. What we precisely mean by nontrivial roots relates to the fact that $0$ is a root of the first derivative of all Bessel's function of the first kind for all integer orders. However, $J_{n_{\scriptscriptstyle L}}(0) \neq 0$ only for $n_{\scriptscriptstyle L} = 0$. Hence $0$ is a non-trivial root of ${J}'_{n_{\scriptscriptstyle L}}(\alpha)$ {\em i.e.}, it is associated with a non-zero eigen function, only for $n_{\scriptscriptstyle L} = 0$. This particular solution corresponds to the (uniform) Kittel's mode. In addition, all the non-zero roots of Eq.~(\ref{eq:exch_sec}) for any value of $n_{\scriptscriptstyle L}$ are obviously nontrivial. We provide some high accuracy (16 digits) approximate values for the $\alpha_{n_R, n_{\scriptscriptstyle L}}$ roots, for $|n_{\scriptscriptstyle L}| \leq 5$ and $n_R \leq 3$, in Table~(\ref{tab:roots_lk}). One can also easily demonstrate, with the help of the solution of problem 3.8 in Ref.~\cite{jackson1962}, that the following particular choice of normalization constant
\begin{equation}
	A_{n_R, n_{\scriptscriptstyle L}} = \frac{\sqrt{2}}{\left[ 1 - \frac{n_{\scriptscriptstyle L}^2}{\alpha_{n_R, n_{\scriptscriptstyle L}}^2} \right]^{1/2} J_{n_{\scriptscriptstyle L}}\left[\alpha_{n_R,n_{\scriptscriptstyle L}} \right]} 
\end{equation}
insures orthonormality among the exchange SWs in the Hermitian \smash{${\rm L}^2$} norm {\em i.e.},
\begin{equation}
    \langle \tilde{\bm m}_{n_R, n_{\scriptscriptstyle L}, n_{\scriptscriptstyle S}}^{(0)} | \tilde{\bm m}_{n_R', n_{\scriptscriptstyle L}', n_{\scriptscriptstyle S}'}^{(0)} \rangle =  \delta_{n_R, n_R'}  \delta_{n_{\scriptscriptstyle L}, n_{\scriptscriptstyle L}'}  \delta_{n_{\scriptscriptstyle S}, n_{\scriptscriptstyle S}'} .  \label{eq:orthonorm1}
\end{equation}
It is also immediate, and relevant, to verify that the orthonormality in  the Hermitian \smash{${\rm L}^2$} norm implies orthonormality in the sense of Eq.~(\ref{eq:sw_ortho}) for circularly polarized azimuthal SWs of the type defined by Eq.~(\ref{eq:exc_SW21}).  

\section{Matrix elements of the inhomogeneous equilibrium DDI}
\label{sec:ie_matelem} 
From Eqs.~(\ref{eq:inho_oper}) and (\ref{eq:other_matelem}), in combination with Eq.~(\ref{eq:exc_SW21}), we obtain
\begin{align} \label{eq:matelem_ied} 
    &{O}_{{\rm i}, n_{\scriptscriptstyle J}}^{n_R, n_{\scriptscriptstyle S}, n_R', n_{\scriptscriptstyle S}} = n_{\scriptscriptstyle S} A_{n_R, n_{\scriptscriptstyle L}} A_{n_R', n_{\scriptscriptstyle L}} \times \cdots \\
    &\times \int_{0}^{1} \!\!  d \underline{r} \ \underline{r} \ J_{n_{\scriptscriptstyle L}} \left( \alpha_{n_R, n_{\scriptscriptstyle L}} \ \underline{r} \right) \ \delta \underline{N}_z(\underline{r}) J_{n_{\scriptscriptstyle L}} \left( \alpha_{n_R', n_{\scriptscriptstyle L}} \ \underline{r} \right) , \nonumber 
\end{align}
with \smash{$n_{\scriptscriptstyle L} = n_{\scriptscriptstyle J} - n_{\scriptscriptstyle S}$}. Numerically efficient and accurate approximations for \smash{$\delta N_z(\underline{r})$}, for any value of the aspect ratio, can be obtained by numerical integration over $\xi$ from Eq.~(\ref{eq:delta_nz}). Once such approximation is obtained, it is straightforward to evaluate the matrix elements as expressed in Eq.~(\ref{eq:matelem_ied}), by numerical integration over $\underline{r}$, for any values of $n_{\scriptscriptstyle J}, n_{\scriptscriptstyle S}, n_R$ and $ n_R'$.

\section{Matrix elements of the dynamical DDI}
\label{sec:ddi_matelem} 
From Eq.~(\ref{eq:ddi_oper}) we immediately have
\begin{equation} \label{eq:matelem_dd1}
    \!\!\!\langle \tilde{\bm m} | \hat{\mathcal O}_{\rm d} | \tilde{\bm m}' \rangle = {\bf i} \!\! \int \!\! d^2 \underline{r} ( {\bf e}_z \times \tilde{\bm m} )^{\star} \!\cdot  {\bm \nabla}_{\underline{r}} \!\! \int \!\! d^2\underline{r}' \tilde{\bm m}' \cdot {\bm \nabla}_{\underline{r}'}  \underline{\rm G}(\underline{\bm r}, \underline{\bm r}') .
\end{equation}
We also recall that one of the valid representation in cylindrical coordinates of the Green's function of the Laplacian, solution of Eqs.~(\ref{eq:poisson_gf})-(\ref{eq:asymptotic_bc_gf}), is given by~\cite{jackson1962}
\begin{align}
    {\rm G}({\bm x},{\bm x}') &= - \frac{1}{4 \pi} \sum_{{n}_J \in {\mathbb Z}} e^{{\bf i} {n}_J (\theta - \theta')} \times \cdots \nonumber \\
    &\times \int_{0}^{+ \infty} \!\!d \xi \  J_{{n}_J} (\xi r) J_{{n}_J} (\xi r') \ e^{- \xi |z - z'|} \ . \label{eq:green_cylind}
\end{align} 
Then, from Eq.~(\ref{eq:perp_gf_nd}), we immediately obtain:
\begin{align}
    &\underline{\rm G}(\underline{\bf r}, \underline{\bf r}') = - \frac{1}{2 \pi} \sum_{{n}_J \in {\mathbb Z}} e^{{\bf i} {n}_J (\theta - \theta')} \times \cdots \nonumber \\
    &\times \int_{0}^{+ \infty} \!\!d \alpha \  \frac{P(\rho \alpha)}{\alpha}  J_{{n}_J} (\alpha \underline{r}) J_{{n}_J} (\alpha \underline{r}') , \label{eq:green_ortho1} 
\end{align}
in which
\begin{equation}
    P(\rho \alpha) = \frac{(\rho \alpha - 1) + e^{-\rho \alpha}}{\rho \alpha} . \label{eq:green_ortho2}
\end{equation}  
It is also useful to define the functions
\begin{align}
    &\ \ \ \ \ \ \ \ \ \ \ \ \ F_{n_{\scriptscriptstyle J}}(\alpha_{0}, \alpha_{1}) = \int_{0}^{1} \!\! d \underline{r} \ \underline{r} J_{n_{\scriptscriptstyle J}} (\alpha_{0} \underline{r})  J_{n_{\scriptscriptstyle J}} (\alpha_{1} \underline{r}) \nonumber \\ 
    &=  \frac{\alpha_{1} J_{|n_{\scriptscriptstyle J}|-1}(\alpha_{1}) J_{|n_{\scriptscriptstyle J}|}(\alpha_{0}) - \alpha_{0} J_{|n_{\scriptscriptstyle J}|-1}(\alpha_{0}) J_{|n_{\scriptscriptstyle J}|}(\alpha_{1}) }{\alpha_{0}^2 - \alpha_{1}^2}. \label{eq:ddi_F}
\end{align}
Then, from Eqs.~(\ref{eq:matelem_dd1}) and (\ref{eq:other_matelem}), in combination with Eqs.~(\ref{eq:exc_SW21}) and (\ref{eq:green_ortho1})-(\ref{eq:ddi_F}), and after some straightforward calculations, one can finally establish 
\begin{widetext}
\begin{equation}
    {O}_{{\rm d}, n_{\scriptscriptstyle J}}^{n_R, n_{\scriptscriptstyle S}, n_R', n_{\scriptscriptstyle S}'} = n_{\scriptscriptstyle S} \frac{A_{n_R, n_{\scriptscriptstyle J} - n_{\scriptscriptstyle S}} A_{n_R', n_{\scriptscriptstyle J} - n_{\scriptscriptstyle S}'}}{2}  \int_{0}^{+ \infty} \!\!\!\!\!\! d \alpha \ \alpha P(\rho \alpha) F_{n_{\scriptscriptstyle J} - n_{\scriptscriptstyle S}}\left[\alpha, \alpha_{n_R, n_{\scriptscriptstyle J} - n_{\scriptscriptstyle S}}\right] F_{n_{\scriptscriptstyle J} - n_{\scriptscriptstyle S}'}\left[\alpha, \alpha_{n_R', n_{\scriptscriptstyle J} - n_{\scriptscriptstyle S}'}\right] .  \label{eq:matelem_ddi} 
\end{equation}
\end{widetext}
The remaining one-dimensional integrals over $\alpha$ in the above expressions can be evaluated numerically. However, one shall be aware that the oscillatory character of the integrands, combined with the unbounded domain of integration, makes accurate numerical evaluations of these matrix elements somewhat challenging. It is nevertheless possible to get reliable numerical approximate values by using advanced algorithms, such as global adaptive schemes combined with suitable quadrature rules.

%

\end{document}